\documentclass{aa}
\usepackage{psfig}
\usepackage{natbib}
\usepackage{longtable}
\usepackage{graphicx}
\bibpunct{(}{)}{,}{a}{}{,} 

\def\la{\;
\raise0.3ex\hbox{$<$\kern-0.75em\raise-1.1ex\hbox{$\sim$}}\; }
\def\ga{\;
\raise0.3ex\hbox{$>$\kern-0.75em\raise-1.1ex\hbox{$\sim$}}\; }

\newcommand{\kms}{km~s$^{-1}$}

\newcommand{\ms}{m~s$^{-1}$}
\newcommand{\cm}{cm$^{-2}$}
\newcommand{\cmm}{cm$^{-3}$}
\newcommand{\etal}{{et al.}}
\newcommand{\nhhh}{NH$_3$}

\newcommand{\Dv}{$\Delta v$}
\newcommand{\Tmb}{$T_{\scriptscriptstyle \rm MB}$}
\newcommand{\VLSR}{$V_{\scriptscriptstyle \rm LSR}$}
\newcommand{\Tkin}{$T_{\rm kin}$}
\newcommand{\Trot}{$T_{\rm rot}$}
\newcommand{\Tex}{$T_{\rm ex}$}

\begin{document}

\title{
Star-forming regions of the Aquila rift cloud complex. 
}
\subtitle{I. NH$_3$ tracers of dense molecular cores}
\author{
S. A. Levshakov\inst{1,2,3}
\and
C. Henkel\inst{4,5}
\and
D. Reimers\inst{1}
\and
M. Wang\inst{6}
\and
R. Mao\inst{6}
\and
H. Wang\inst{6}
\and
Y. Xu\inst{6}
}
\institute{
Hamburger Sternwarte, Universit\"at Hamburg,
Gojenbergsweg 112, D-21029 Hamburg, Germany
\and
Ioffe Physical-Technical Institute,
Polytekhnicheskaya Str. 26, 194021 St.~Petersburg, Russia
\and
St.~Petersburg Electrotechnical University `LETI', Prof. Popov Str. 5,
197376 St.~Petersburg, Russia \\
\email{lev@astro.ioffe.rssi.ru}
\and
Max-Planck-Institut f\"ur Radioastronomie, Auf dem H\"ugel 69, D-53121 Bonn, Germany
\and
Astronomy Department, King Abdulaziz University, P.O.
Box 80203, Jeddah 21589, Saudi Arabia
\and
Purple Mountain Observatory, Academia Sinica, Nanjing 210008, P. R. China
}
\date{Received 00  ; Accepted 00}
\abstract
{}
{The physics of star formation is an important issue of 
Galactic evolution. Most stars are formed in high density environments 
($n > 10^4$ \cmm) emitting lines of diverse molecular transitions. 
In the present part of our survey
we search for ammonia emitters in the Aquila rift complex 
which trace the densest regions of molecular clouds. 
}
{From a CO survey carried out with the Delingha 14-m telescope we selected
$\sim 150$ targets for observations in other molecular lines.
Here we describe the mapping observations in the \nhhh(1,1) and (2,2) 
inversion lines of the first 49 sources performed with the Effelsberg 100-m telescope. 
}
{The \nhhh(1,1) and (2,2) emission lines are detected in 12 and 7 sources, respectively.
Among the newly discovered \nhhh\ sources,
our sample includes the following well-known clouds: the starless core \object{L694-2},
the \object{Serpens cloud Cluster B}, the \object{Serpens dark cloud} \object{L572},
the filamentary dark cloud \object{L673}, the isolated protostellar source \object{B335},
and the complex star-forming region \object{Serpens South}. 
Angular sizes between $40''$ and $80''$ ($\sim 0.04-0.08$ pc)
are observed for compact starless cores but as large as $9'$ ($\sim 0.5$ pc)
for filamentary dark clouds. 
The measured kinetic temperatures of the clouds lie between 9~K and 18~K. 
From \nhhh\ excitation temperatures of $3-8$~K we determine H$_2$ densities
with typical values of $\sim (0.4-4)\times10^{4}$ \cmm. The masses of the mapped cores
range between $\sim 0.05$ and $\sim 0.5M_\odot$. The relative ammonia
abundance, $X=$ [\nhhh]/[H$_2$], varies from $1\times10^{-7}$ to $5\times10^{-7}$
with the mean $\langle X\rangle = (2.7\pm0.6)\times10^{-7}$
(estimated from spatially resolved cores assuming the filling factor $\eta = 1$). 
In two clouds, we observe kinematically split \nhhh\ profiles separated
by $\sim 1$ \kms. The splitting is most likely due to bipolar molecular
outflows for one of which we determine an acceleration of
$\dot{V} \la 0.03$ km~s$^{-1}$~yr$^{-1}$. 
A starless core with significant rotational energy is found to have 
a higher kinetic temperature than the other ones which is probably caused by 
magnetic energy dissipation.
}
{}

\keywords{ISM: clouds --- ISM: molecules --- ISM: kinematics and dynamics ---
Radio lines: ISM --- Line: profiles --- Techniques: spectroscopic
} 

\authorrunning{S. A. Levshakov \etal\ }

\titlerunning{\nhhh\ observations in the Aquila rift cloud complex}

\maketitle

\section{Introduction}
\label{sect-1}

One of the central questions of stellar evolution is about the processes defining 
the initial mass of a protostellar object.
It is suggested that more than 70\% of stars are formed in clusters embedded within giant molecular clouds
(Lada \& Lada 2003) and that the varying environment may affect the initial mass function (IMF).

In the present survey we search for the star-formation regions in the Aquila giant molecular cloud.
Starting from Cygnus, in the direction of the Galactic center, the Milky Way appears to be split into
two branches. The southern branch runs through Cygnus, Vulpecula, Sagitta, Aquila, and Scutum,
entering the Galactic central bulge in Sagittarius. The northern branch crosses Vulpecula and Aquila
and disappears in the northern part of the Serpens Cauda and Ophiuchus constellations. The separating
dark band of obscuring dust is commonly called the Aquila rift, but absorbing clouds are found not only
next to the Galactic plane: many dark clouds are located in areas 
extending up to latitudes of at least +10$^\circ$ and down to 
$-20^\circ$ (e.g., Dame \& Thaddeus 1985; Kawamura \etal\ 1999; Dobashi \etal\ 2005). 

The distance to the Aquila rift was estimated from optical data and interstellar extinction,
and from parallax measurements. 
A sudden appearance of reddened stars is observed at $D = 225\pm55$ pc and
the cloud distribution may have a thickness of $\sim80$ pc 
(Strai$\check{\rm z}$ys \etal\ 2003). This estimate is in line with the distance to the Serpens-Aquila region
found from interstellar extinction by Knude (2011), 
$D = 203\pm7$ pc, but significantly smaller than the trigonometric parallax value $415\pm25$ pc deduced
by Dzib \etal\ (2010) from the Very Long Baseline Array observations 
of the young stellar object \object{EC 95} in the Serpens cloud. 
Since the discrepancy in the distance estimations may be caused by a projection effect,
we will use hereafter the weighted harmonic mean of the first two measurements, both
based on large-size data sets, $D = 203\pm12$ pc, if not otherwise specified.   
Thus, comparing to other close molecular complexes,
the distance to the Aquila rift is similar to that of the Perseus clouds ($\sim 250$ pc; Rosolowsky \etal\ 2008) 
and lies between those of
the Taurus and the Pipe Nebula ($\sim 140$ pc; Rathborne \etal\ 2008)
and the Auriga ($\sim 300$ pc) molecular clouds (Heiderman \etal\ 2010; Cieza \etal\ 2012). 

Recently the {\it Herschel} Gould Belt Survey reveals in
the Aquila rift cloud complex a large number ($\ga 500$)
of starless cores (e.g., Bontemps \etal\ 2010; K\"onyves \etal\ 2010; Men'shchikov \etal\ 2010). 
Starting our investigations of the Aquila clouds
we have carried out in 2010 an extended survey of CO and extinction peaks 
using the Delingha 14-m telescope (Zuo \etal\ 2004).
The targets were selected from 
the CO data of Kawamura \etal\ (1999, 2001) and from the optical
data of Dobashi \etal\ (2005), limited in the latter case by the areas with $A_v > 6^{\rm m}$. 
We observed and detected 
23 Kawamura and 47 Dobashi clouds at $\lambda \sim 3$~mm. 
However, CO and $^{13}$CO trace not only the dense gas but also outflows and low density regions; 
in addition, CO may be saturation broadened. 
It is well known that the carbon-chain molecules disappear from the gas-phase in the central regions
because of freeze-out onto dust grains (e.g., Tafalla \etal\ 2004), i.e.,
C-bearing molecules are usually distributed in the outer parts of the cores. 
That is why CO and $^{13}$CO are not good indicators for the dense molecular gas. 
Contrary to this, 
N-bearing molecules such as ammonia concentrate in the inner cores, where the gas density
approaches $10^5$ \cmm. Ammonia is still observable in the gas-phase since it is resistant
to depletion onto dust grains (e.g., Bergin \& Langer 1997).

Based on the CO survey we composed a list of $\sim 150$ targets where
not only CO but also $^{13}$CO is quite strong, thus indicating large column densities
and hence promising conditions to detect dense cores. 
The objective of the current work
performed with the Effelsberg 100-m telescope is to observe these targets in
other molecular lines, in the first place in \nhhh, in order to confirm the
target identifications as dense cores and to determine their physical properties.

In the present paper we describe the first 49 sources from our survey.
Observations are described in Sect.~\ref{sect-2}.
Individual molecular cores are considered in Sect.~\ref{sect-4}.
The results obtained are summarized in Sect.~\ref{sect-5}, and 
computational methods are outlined in Appendix A. 
Appendix B contains \nhhh(1,1) and (2,2) spectra detected toward the ammonia peaks, tables with
derived physical parameters, and maps showing distributions of these parameters within the three
most prominent cores.

\section{Observations}
\label{sect-2}

The ammonia observations were carried out with the Effelsberg 100-m
telescope\footnote{The 100-m telescope at Effelsberg/Germany is operated
by the Max-Planck-Institut f{\"u}r Radioastronomie on behalf of the
Max-Planck-Gesellschaft (MPG).}
in 4 observing sessions between March 26 and 30, 2011, and in 2 sessions
on January 12 and 13, 2012.
The \nhhh\ lines were
measured with a K-band HEMT (high electron mobility transistor)
dual channel receiver, yielding spectra with full width to half power (FWHP) resolution of $40''$ 
in two orthogonally oriented linear polarizations 
at the rest frequency of the $(J,K) = (1,1)$ and (2,2) lines
(23694.496 MHz and 23722.631 MHz, respectively).
Averaging the emission from both
channels, the typical system temperature (receiver noise and atmosphere)
is 100\,K on a main beam brightness temperature scale.

In 2011, the measurements were obtained in frequency switching mode
using a frequency throw of $\pm 2.5$\,MHz. The backend was an FFTS
(fast Fourier transform spectrometer), 
operated with a bandwidth of 100\,MHz, providing simultaneously 16\,384 channels
for each polarization. For some offsets we also used the minimum
bandwidth of 20\,MHz for the \nhhh(1,1) line.
The resulting channel widths were 0.077 \kms\ and
0.015 \kms, respectively. We note,
however, that the true velocity resolution was about 1.6 times as large.

In 2012, the measurements were obtained 
with the backend XFFTS (eXtended bandwidth FFTS), 
operated with 100\,MHz and 2\,GHz bandwidths, providing 32\,768 channels
for each polarization. The resulting channel widths were 0.038 \kms\ and 0.772 \kms, 
respectively, but the true velocity resolution is 1.16 times larger (Klein \etal\ 2012).

The pointing was checked every hour by continuum cross scans of nearby
continuum sources. The pointing accuracy was better than $5''$. 
The spectral line data were calibrated by means of continuum sources
with known flux density. We mainly used NGC7027 (Ott \etal\ 1994),
3C123 (Baars \etal\ 1977), and G29.96--0.02 (Churchwell \etal\ 1990).
These calibration sources were used to establish a main beam brightness
temperature scale, \Tmb. 
Since the main beam size ($40''$) is smaller than most core radii ($> 50''$)  
of our Aquila objects, the ammonia
emission couples well to the main beam and, thus, the \Tmb\ scale is
appropriate. 
Compensations for differences in elevation
between the calibrator and the source were 
$\la 20$\% and have not been taken into account.
Similar uncertainties of the main beam brightness temperature 
were found from comparison of spectra toward the same position taken at different dates.

\section{Results}
\label{sect-4}

The detected \nhhh\ spectra were treated in the way described in Appendix~A.
We note that 
the radial velocity, $V_{\scriptscriptstyle \rm LSR}$,  the linewidth  \Dv\ (FWHM), the  
optical depths, $\tau_{11}$ and $\tau_{22}$, the integrated ammonia emission, $\int$\Tmb$dv$, and
the kinetic temperature, \Tkin, are well determined physical parameters, whereas
the excitation temperature, \Tex, the ammonia column density, $N$(\nhhh), and the gas density,
$n({\rm H}_2)$ are less certain.
The optical depth and kinetic temperature 
are parameters independent of calibration errors, since \Tkin\ 
depends only on the intensity ratio of the simultaneously
observed \nhhh(1,1) and (2,2) lines, and $\tau$ depends
on the relative intensities of the various hyperfine (hf) components. 
On the other hand,
the excitation temperature \Tex\ depends on the ammonia
line brightness and is sensitive to calibration errors (Lemme \etal\ 1996).
Besides, Eq.(\ref{Eq11}) shows that \Tex\ depends on the filling factor $\eta$
which is not known for unresolved cores. 

A typical error of the optical depth determined formally 
from the covariance matrix at the minimum of $\chi^2$ is 15\%-20\% (Appendix A). 
The error of the integrated ammonia emission depends
on the noise level and for an rms $\sim 0.2$~K per 0.077 \kms\ channel 
it is $\sim 0.3$ K~\kms.
The precision with which the Gaussian line center can be estimated
is given by (e.g., Landman et al. 1982)
\begin{equation}
\sigma_V \approx \frac{1}{(2\pi \ln{2})^{1/4}} \frac{rms}{T_{\scriptscriptstyle \rm MB}} 
\sqrt{\Delta_{\rm ch} \Delta v}\, ,
\label{SigV}
\end{equation}
where $\Delta_{\rm ch}$ and $\Delta v$ are the channel and the line widths, respectively.
At rms $\sim 0.2$~K and \Tmb $\sim 2$~K (see below) it gives $\sigma_V \sim 0.005$ \kms\
for our high resolution settings. The same order of magnitude errors (i.e., from a few to ten \ms)
of the line centers and linewidths are obtained from the covariance matrix.

The results of our analysis are presented in Tables~\ref{tbl-1}-\ref{tbl-2} and \ref{tbl-3}-\ref{tbl-6}.
Table~\ref{tbl-1} lists the regions where \nhhh\ line emission was detected either in both (1,1) and
(2,2) transitions or only in (1,1).
In case of non-detections (Table~\ref{tbl-2}), the rms \Tmb\ values per channel width are given. 
In total, \nhhh\ lines are detected in 12 sources out of 49 dense cores.
In Table~\ref{tbl-1}, we list 
the radial velocities, $V_{\scriptscriptstyle \rm LSR}$,  the observed linewidths  \Dv\ (FWHM),
and the main-beam-brightness temperature,  
$T_{\scriptscriptstyle\rm MB}$, at the maximum of the main group of hf components.
In Col.~2, different peaks resolved in the sources are labeled by Greek letters ($\alpha$ for
the most intense \nhhh\ emission) which are also indicated on color maps 
(Figs.~\ref{Fg1}-\ref{Fg3}, \ref{Fg5}, and \ref{Fg7}-\ref{Fg10}). 
The physical parameters calculated for each individual source are given in Table~\ref{tbl-1} and in
the Appendix~B, Tables~\ref{tbl-3}--\ref{tbl-6}. The results obtained for the three most prominent sources 
\object{Do279P6}, \object{Do279P12}, and \object{SS3} are shown in addition as maps (Figs.~\ref{afg10}-\ref{afg12}).
Below we describe the individual sources identified as dense molecular cores. 
Statistical analysis of the full sample of Aquila cores will be reported after the survey
will be completed in 2013.

\subsection{\object{Kawamura~01} (\object{Ka01})}
\label{sect-4-1}

The starless cloud \object{Kawamura~01} (\object{Ka01} herein) 
belongs to the dark cloud \object{L694}  
classified as opacity class 6 by Lynds (1962).
This is an isolated dense
core located in the sky close ($\sim 4^\circ$) to the protostar B335
(Sect.~\ref{sect-4-6}).
No embedded {\it IRAS} luminosity source has been found 
(Harvey \etal\ 2002; Harvey \etal\ 2003a).
Lee \etal\ (1999) found three dark spots \object{L694--1}, \object{L694--2}, and \object{L694--3}
in the Digital Sky Survey image of \object{L694}. 
The second spot is associated with \object{Ka01}. 
Using Wolf diagrams, Kawamura \etal\ (2001) determined the
distance to \object{L694-2} as $D \approx 230$ pc.

\begin{figure}[t]
\vspace{0.0cm}
\hspace{-1.0cm}\psfig{figure=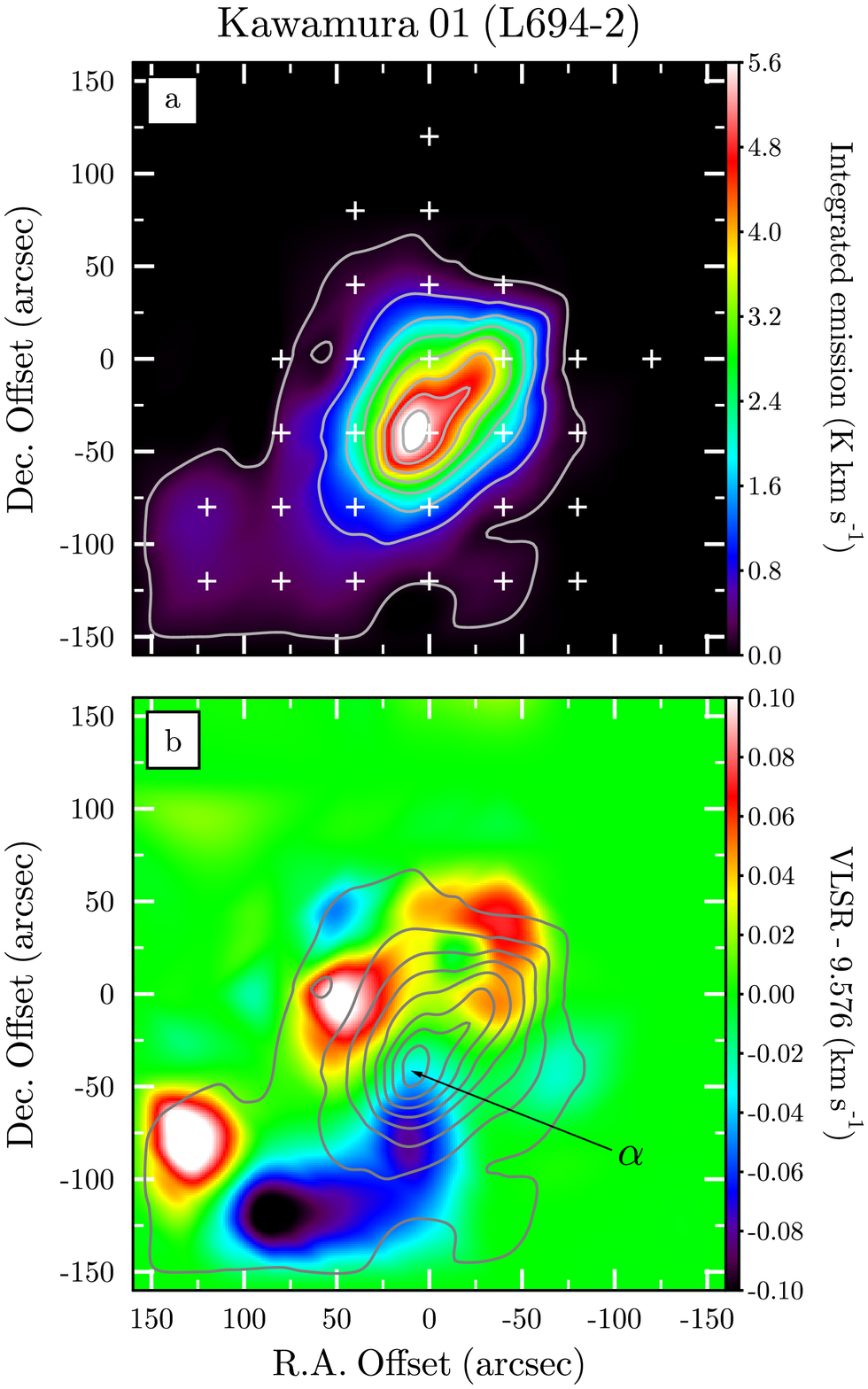,height=14.0cm,width=11.0cm}
\vspace{-0.7cm}
\caption[]{
({\bf a}): \nhhh(1,1) intensity map ($\int$\Tmb$dv$) of \object{Ka01}. 
The starting point for the contour levels is 0.1 K~\kms, the increment
is 0.7 K~\kms\ between the first two and 0.8 K~\kms\ between the other contour 
levels. The crosses mark measured positions. 
({\bf b}): \nhhh(1,1) radial velocity (color map) structure. 
The velocity field is shown after subtraction of the mean radial
velocity $\langle V_{\scriptscriptstyle\rm LSR} \rangle = 9.576$ \kms.
The intensity peak is labeled in accord with Table~\ref{tbl-1}. 
The Effelsberg beam size (HPBW) at 23.7 GHz is $40''$.
}
\label{Fg1}
\end{figure}

We mapped \object{Ka01} for the first time
in the \nhhh(1,1) and (2,2) lines with a spacing of $40''$ 
at 29 positions marked by crosses in Fig.~\ref{Fg1}{\bf a}. 
It is seen that the \nhhh\ emission arises from a very compact region. 
The fourth contour level in Fig.~\ref{Fg1}{\bf a}
corresponds to the half-peak of the integrated emission, $\int${\Tmb}$dv$.
This allows us to calculate the apparent geometrical mean diameter (FWHP) 
$d = \sqrt{ab}$, where $a \approx 115''$ and $b \approx 65''$ (deconvolved)
are the major and minor axis of this annulus, respectively,
and a beam ammonia core size of approximately 0.08 pc~---  the typical value 
for dense molecular cores in the Taurus molecular cloud (e.g., Benson \& Myers 1989). 

At the core center,
we obtained two spectra with low (FWHM = 0.123 \kms) and high (FWHM = 0.024 \kms)
spectral resolutions. 
The high resolution spectra of the
\nhhh(1,1) and (2,2) lines are shown in Fig.~\ref{afg1} along with 
the synthetic spectra (red curves) calculated in the
simultaneous fit of all hf components to the observed profiles (blue histograms). 
The residuals ``observed data -- model'' are depicted beneath each \nhhh\ spectrum in
Fig.~\ref{afg1}. 

The results of the fits to the low-resolution \nhhh\ spectra obtained at different
positions and the corresponding physical
parameters are given in Table~\ref{tbl-3}.
The estimate of the excitation temperature is made from the (1,1) 
transition assuming a beam filling factor $\eta = 1$ 
since both the major and minor axes of the core
exceed the angular resolution. 
At two offsets $(0'',0'')$ and $(0'',-40'')$, 
we calculated the rotational and kinetic temperatures 
from the two transitions (1,1) and (2,2), and
from Eq.(\ref{Eq19}) we determine the gas densities $n_{{\rm H}_2} \sim 1.2\times10^4$ and $2.5\times10^4$ \cmm,
respectively, for a uniform cloud coverage (Col.~10 in Table~\ref{tbl-3}).

Assuming spherical geometry, we find a core mass 
$M \sim 0.5M_\odot$ (the mean molecular weight is 2.8). 
On the other hand, the virial mass [Eq.(\ref{Eq23})] calculated from
the linewidth $\Delta v$ and the core radius $r$ is
$M_{\rm vir} \sim 1.0M_\odot$. The observed difference in the masses could be due to deviations from a uniform
gas density distribution and a core ellipticity. 

The total ammonia column density of $N_{\rm tot} \sim 1.3\times10^{15}$ \cm\
gives the abundance ratio [NH$_3$]/[H$_2$]~$\sim 2\times10^{-7}$ at the
core center ($d \sim 0.08$ pc) which is consistent with other sources
(e.g., Dunham \etal\ 2011). 

The high resolution \nhhh\ spectrum (Fig.~\ref{afg1})
can be used to estimate the input of the turbulent motions to the line broadening.
The measured linewidth \Dv\ = 0.27 \kms\
and the thermal width $\Delta v_{\rm th} \sim 0.15$ \kms\
at \Tkin\ $\sim 9$~K (Table~\ref{tbl-3}) give 
$\sigma_{\rm turb} \sim 0.09$ \kms.

Figure~\ref{Fg1}{\bf b} shows the
\nhhh(1,1) intensity (grey contours) and the radial velocity $V_{\scriptscriptstyle \rm LSR}$ (color map) 
structure in \object{L694-2}. 
The radial velocities within the central $115'' \times 65''$ region
do not show significant variation in both magnitude and direction, 
suggesting a simple solid-body rotation. The velocity gradient along the major axis (P.A. = $127^{\circ}$)
is ${\rm grad}_a V \equiv (V - \langle V_{\scriptscriptstyle \rm LSR} \rangle)/0.5a 
\sim \pm 1$ km~s$^{-1}$~pc$^{-1}$, and along the minor axis 
${\rm grad}_b V \equiv (V - \langle V_{\scriptscriptstyle \rm LSR}\rangle)/0.5b \sim 0$ and $+1$ km~s$^{-1}$~pc$^{-1}$. 
Towards the edges N-W (redshift) and S-E (blueshift), $\sim 50''$ off the center, 
the radial velocity monotonically changes by $\pm0.06$ \kms.  
Being attributed to cloud rotation, such systematic trends in \VLSR\ 
correspond to an angular
velocity $\dot{\phi} \approx 4\times10^{-14}$ s$^{-1}$.   

For a self-gravitating rigidly rotating sphere of constant density $\rho$, mass $M$, and radius $r$,
the ratio of rotational to gravitational potential energy is
\begin{equation}
\beta = \left| \frac{E_{\rm rot}}{U} \right| = \frac{3 M {\dot{\phi}}^2 }{16 \pi^2 G \rho^2 r^3} =
\frac{ {\dot{\phi}}^2 }{4 \pi G \rho} \, ,
\label{beta1}
\end{equation}
or
\begin{equation}
\beta = 2.55\times10^{-3} \ {\dot{\phi}}^2_{-14}/n_4 \, ,
\label{beta2}
\end{equation}
where ${\dot{\phi}}_{-14}$ is the angular velocity in units of $10^{-14}$ s$^{-1}$ and $n_4$ is the gas
density in units of $10^4$ \cmm\ (Menten \etal\ 1984).

Equation~(\ref{beta2}) and the estimates of the angular velocity and the gas density 
obtained above show that $\beta \la 0.016$ and, hence, 
the rotation energy is a negligible fraction of the gravitational energy at this stage of evolution 
of the low mass starless molecular core \object{Ka01}.  

There is a different measure of the influence of cloud rotation upon cloud stability
through the ratio between the rotational and combined thermal and non-thermal (turbulent)
virial terms (e.g., Phillips 1999):
\begin{equation}
\beta' = 7.04\times10^{-2}\ r^2\ {\dot{\phi}}^2_{-14}\ {\Delta v}^{-2}\ , 
\label{beta3}
\end{equation}
where $r$ is the cloud radius in pc, and $\Delta v$ the linewidth (FWHM) in \kms.
The influence of rotation and turbulence are comparable when 
the stability parameter $\beta'(r) \approx 1$.
For the core \object{Ka01}, $\beta' \approx 0.02$ which means that turbulence 
greatly exceeds the contribution due to rotation in determining cloud stability.

\subsection{\object{Dobashi~279~P6} (\object{Do279~P6})}
\label{sect-4-2}

\begin{figure}
\vspace{0.0cm}
\hspace{-3.5cm}\psfig{figure=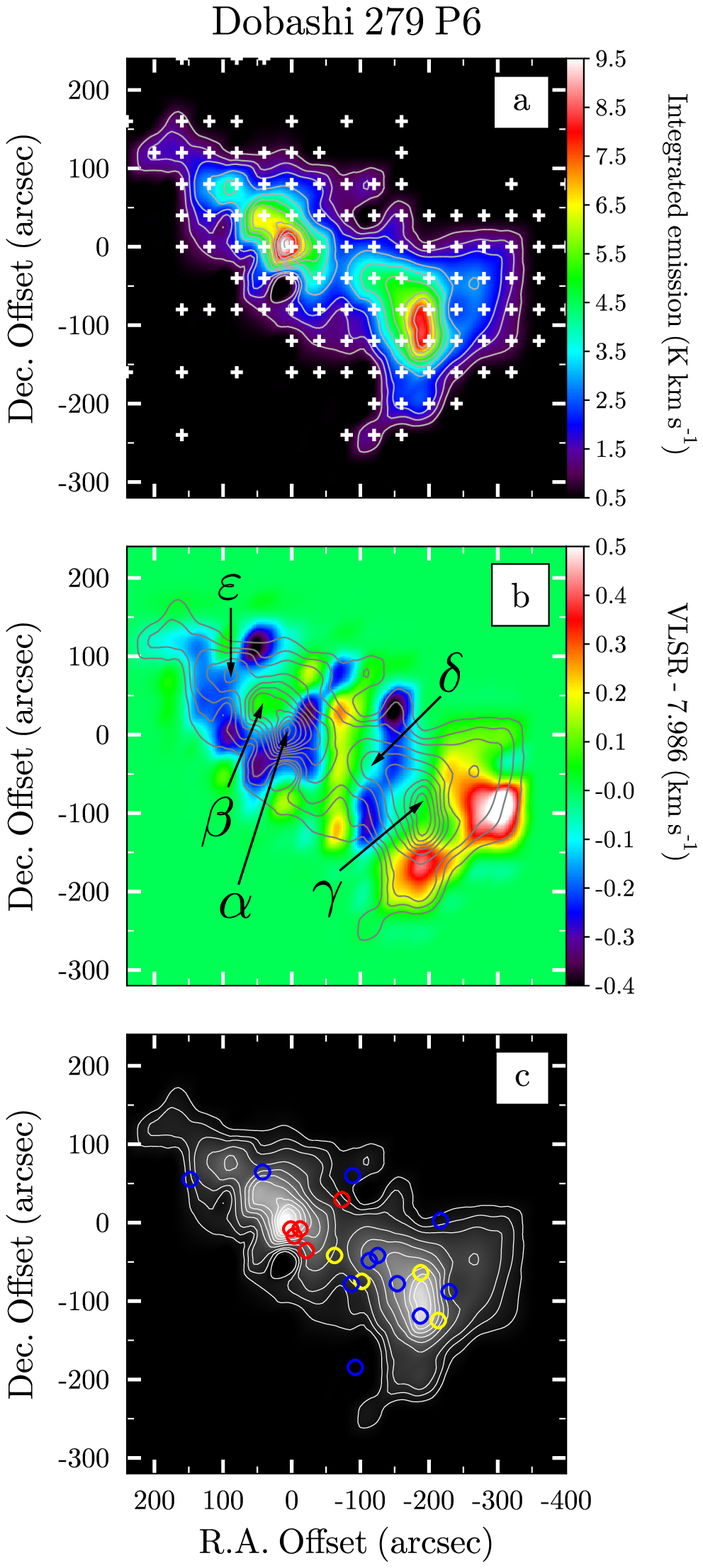,height=20.0cm,width=18.0cm}
\vspace{-3.0cm}
\caption[]{
({\bf a}): Same as Fig.~\ref{Fg1}{\bf a} but for the source \object{Do279P6}. 
The starting point for the contour levels is 0.5 K~\kms, the increment
is 0.5 K~\kms\ between the first two and 1.0 K~\kms\ between the other contour levels.
({\bf b}): \nhhh(1,1) radial velocity (color map) structure.
The velocity field is shown after subtraction of the mean radial
velocity $\langle V_{\scriptscriptstyle\rm LSR} \rangle = 7.986$ \kms.
The intensity peaks are labeled in accord with Table~\ref{tbl-1}. 
({\bf c}): \nhhh(1,1) intensity map as in panel {\bf a}
and positions of YSO candidates  
observed with the {\it Spitzer}/IRAC in a star formation region 
Serpens cluster B  (Harvey \etal\ 2006). 
Open circles of different colors mark 
Class~I (red), intermediate type (yellow), and Class~II (blue) YSOs.
}
\label{Fg2}
\end{figure}

The source \object{Dobashi~279~P6} (\object{Do279P6}, herein)
shows a very complex structure in the \nhhh\ emission (Fig.~\ref{Fg2}{\bf a})
with many small clumps located along the main axis (P.A.= $56^\circ$,
$a \approx 9'$, or $\sim 0.5$ pc at $D \sim 203$ pc). 
We list the coordinates of five bright \nhhh\ peaks,
the peak temperatures \Tmb, their radial velocities $V_{\scriptscriptstyle \rm LSR}$, and 
the linewidths $\Delta v$ in Table~\ref{tbl-1}. The peaks are
labeled from $\alpha$ to $\varepsilon$ in Fig.~\ref{Fg2}{\bf b} and in Table~\ref{tbl-1} 
in accord with decreasing main beam brightness temperature, \Tmb. 

The location of \object{Do279P6} coincides with 
a star formation region in the Serpens cloud, which is known under the name
Cluster B  (Harvey \etal\ 2006), or \object{Ser G3-G6} (Cohen \& Kuhi 1979),
or \object{Serpens NH$_3$} (Djupvik \etal\ 2006). 
The area around Ser~G3-G6 was mapped in the \nhhh(1,1) emission line by Clark (1991)
who found two ammonia condensations on each side of the complex, Ser~G3-G6NE
and Ser~G3-G6SW (the map is not published; observations obtained with the Effelsberg 100-m telescope).
The coordinates of these condensations coincide,
respectively, with the $\alpha$ and $\gamma$ ammonia peaks,
whereas the close group of G3-G6 stars lies in the field of the $\delta$ peak. 

The \object{Do279P6} cloud with identified YSO candidates is
shown in Fig.~\ref{Fg2}{\bf c} in which open circles of different colors mark
Class~I (red), ``flat'' (yellow)\footnote{A ``flat'' category is between Classes I and II 
(Andr\'e \& Montmerle 1994; Greene \etal\ 1994; Deharveng \etal\ 2012).},
and Class~II (blue) candidates. 
Class II sources are more evolved and older than Class~I
and, thus, the \nhhh\ core might be already more dispersed.
So, what we observe at the $S-W$ offset $\gamma = (-160'',-80'')$ is consistent with this 
expectation~-- only ``flat'' and Class~II sources are located therein.
On the other hand, 
all Class~I sources (ID numbers B13 and B15--B18 in accord with Table~4 in Harvey \etal\ 2006)
are detected in the vicinity of the core center $\alpha = (0'',0'')$ 
which has the highest gas density, 
$n_{{\rm H}_2} \sim 2.3\times10^4$ \cmm, 
and the maximum intensity of the ammonia emission \Tmb\ = 2.9~K (Table~\ref{tbl-4}).
We note that the source B18 lies
within $8''$  off the \nhhh\ peak,
and B13~-- the most distant~-- at $\sim 80''$ along the N--W direction. 
It is also noticeable that the angular size of the central condensation is two times
smaller then that of the S--W clump, i.e., the region is probably less dispersed.
The two other condensations with  ``flat'' and Class~II type sources
are located at the offsets $\delta = (-120'',-40'')$ and $\beta = (40'',40'')$ with, respectively,
five and one embedded YSO(s).
The only starless \nhhh\ region detected at the $\varepsilon$ condensation 
$\varepsilon = (80'',80'')$ has
the narrowest linewidth $\Delta V \simeq 0.4$ \kms\ (Table~\ref{tbl-1}) and
the lowest kinetic temperature $T_{\rm kin} \simeq 10$~K (Table~\ref{tbl-4})
which are comparable to \object{Ka01}~--- the isolated dense core (Sect.~\ref{sect-4-1}).

Figure~\ref{afg2} shows the high-resolution ammonia spectra (channel spacing 0.015 \kms)
obtained at three \nhhh\ peaks: $\alpha$, $\beta$, and $\gamma$. 
The corresponding linewidths are the following:
$\Delta v \approx 0.75, 0.63$, and 0.96 \kms. 
Given the derived kinetic temperatures (Table~\ref{tbl-4}),
we can calculate the thermal contribution to the linewidth and the non-thermal velocity component:
\begin{equation}
\Delta v^2_{\scriptscriptstyle \rm NT} = \Delta v^2 - 8\ln 2 \frac{k T_{\rm kin}}{17m_{\rm H}} , 
\label{dispV}
\end{equation}
where $(k T_{\rm kin}/17m_{\rm H})$ is the thermal broadening due to \nhhh, and $m_{\rm H}$ is the mass
of the hydrogen atom.

The calculated non-thermal velocity dispersion,
$\sigma_{\scriptscriptstyle \rm NT} = \Delta v_{\scriptscriptstyle \rm NT}/(2\sqrt{2\ln 2})$, 
of about 0.3 \kms\ is comparable to the large scale 
motions observed in the cloud. 
Namely, the $N-E$ part is moving towards us 
(with respect to the systemic velocity of the cloud,
$\langle V_{\scriptscriptstyle \rm LSR}\rangle = 7.986$ \kms)
with a radial velocity  $V_{\scriptscriptstyle \rm LSR} \approx -0.4$ \kms\ 
and the $S-W$ part in the opposite direction with similar radial velocity (Fig.~\ref{Fg2}{\bf b}).
In the middle of the cloud there is a narrow zone along the cut $\Delta \alpha = -50''$ 
with positive \VLSR\ $\approx 0.3$ \kms, which is bracketed by two zones with negative 
\VLSR\ $\approx -0.3$ \kms. 
The whole structure may be due to the differential rotation of different clumps distributed along the
main axis of the cloud. 

The central ammonia condensation consists of two clumps $\alpha$ and $\beta$
separated by $\sim 60''$. The position angle of the major axis of the brighter $\alpha$ clump 
is P.A. $\sim 50^\circ$. The deconvolved diameters of the major and minor axes are $a \simeq 130''$ and 
$b \simeq 60''$ (FWHP). The angular sizes of the $\beta$ clump are less certain since it is not
completely resolved. The physical parameters were estimated assuming a filling factor $\eta = 1$
for both the $\alpha$ and $\beta$ clumps and thus 
\Tex, $N$(\nhhh), and $n({\rm H}_2)$, listed in Table~\ref{tbl-4}, 
correspond to their minimum values (for details, see Appendix~A).
With the mean diameter of $\sim 90''$ (or 0.09 pc at D = 203 pc) the mass of this double clump is
$M_{\alpha,\beta} \ga 0.5M_\odot$, and the ammonia abundance is 
[\nhhh]/[H$_2$] $\sim 2\times10^{-7}$.
The estimate through virial masses is less certain in this case since YSOs are embedded 
within the clumps. 

Another double peak is formed by the $\gamma$ and $\delta$ cores. The major axis of the
brighter $\gamma$ condensation (P.A. = $0^\circ$) is $113''$ and
the minor axis is $69''$ (deconvolved, FWHP). The $\delta$ clump is not completely resolved
which gives, as for the previous pair, the lower bound on the mass 
$M_{\gamma,\delta} \ga 0.4M_\odot$ and the abundance ratio [\nhhh]/[H$_2$] $\sim 1\times10^{-7}$. 

The weakest peak, $\varepsilon$, separated by $\sim 60''$ from $\beta$, is
a completely unresolved ammonia clump with an angular size $\theta < 40''$.
It shows the largest optical depth, $\tau_{11} \simeq 11$, and the lowest kinetic temperature,
\Tkin $\simeq 10$~K (Table~\ref{tbl-4}). Other physical parameters listed in Table~\ref{tbl-4}
were calculated at $\eta = 1$, i.e., they are the minimum values. The filling factor $\eta$ is restricted
in this case between 0.2 and 1 (see Appendix A), which provides the following boundaries:
$4 \leq T_{\rm ex} < 10$~K, $1.3\times10^{15} \leq N({\rm NH}_3) < 3.2\times10^{15}$ \cm,
$0.8\times10^4 \leq n({\rm H}_2) < 4.9\times10^4$ \cmm.  The mass of the $\varepsilon$ clump is less
than $0.06M_\odot$ if $n({\rm H}_2) \sim 3\times10^4$ \cmm\ and its diameter $d < 0.04$ pc. 

Of course, the derivation of relative ammonia abundances 
and masses is not very accurate and based on a number of
model assumptions like spherical geometry, homogeneous gas density,
the distance to the target of 203~pc, etc.
Nevertheless, the substellar mass of 
the $\varepsilon$ clump may lie
in the range of the lowest possible masses with which 
a brown dwarf can form: $0.03-0.08M_\odot$ (e.g., Oliveira \etal\ 2009).

\subsection{\object{Dobashi~279~P8} (\object{Do279~P8})}
\label{sect-4-3}

\begin{figure}
\vspace{0.0cm}
\hspace{-1.0cm}\psfig{figure=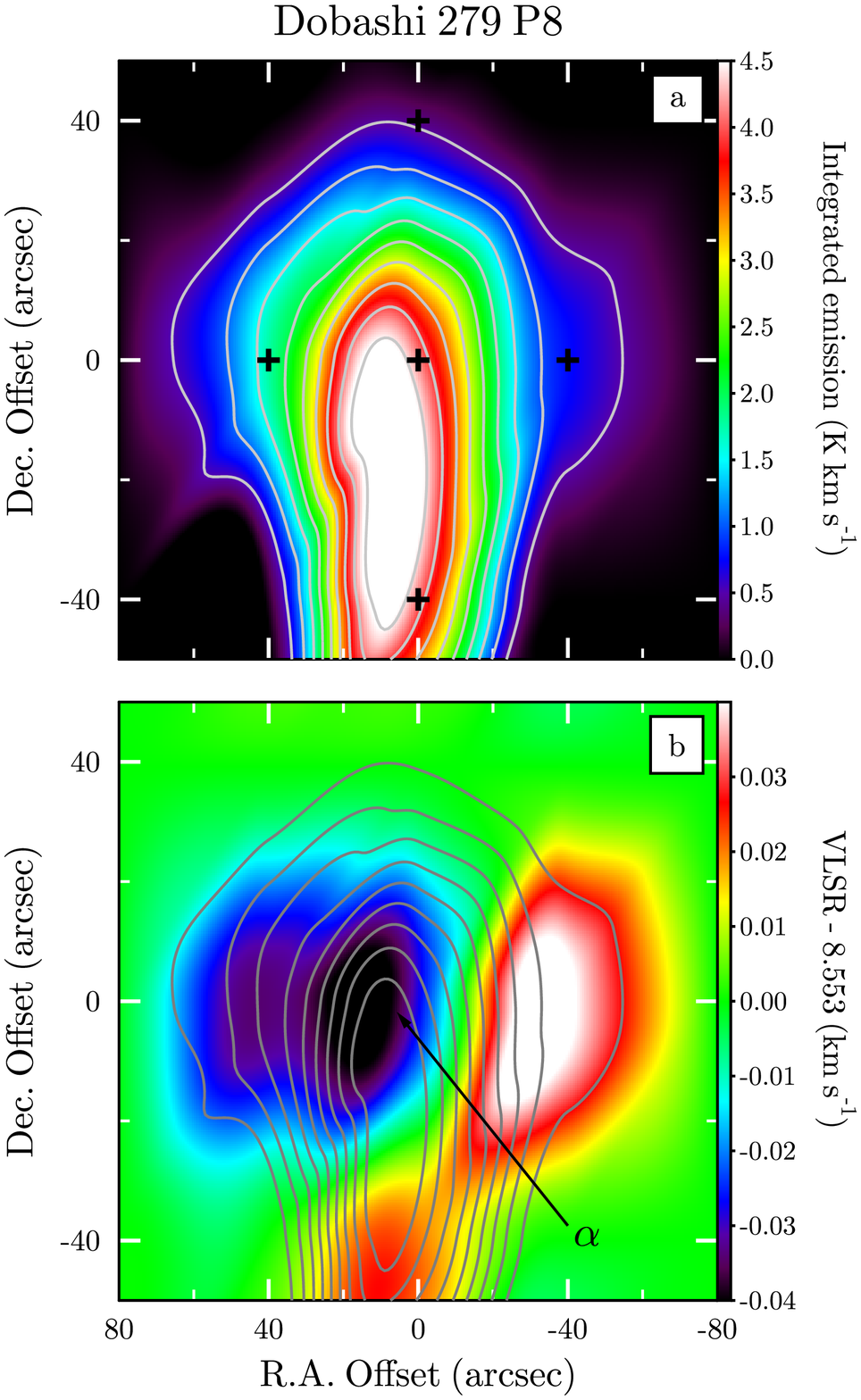,height=14.0cm,width=11.0cm}
\vspace{-0.7cm}
\caption[]{
({\bf a}): Same as Fig.~\ref{Fg1}{\bf a} but for the source \object{Do279P8}. 
The contour levels start at 0.5 K~\kms\ and increments are 0.5 K~\kms. 
({\bf b}):
\nhhh(1,1) radial velocity (color map) structure. 
The velocity field is shown after subtraction of the mean radial
velocity $\langle V_{\scriptscriptstyle\rm LSR} \rangle = 8.553$ \kms.
The intensity peak is labeled in accord with Table~\ref{tbl-1}. 
}
\label{Fg3}
\end{figure}

The molecular cloud \object{Dobashi~279~P8} (\object{Do279P8}, herein)
is an example of a starless core  with no embedded IRAS point sources
or a pre-main-sequence star.
Our mapping reveals a slightly elongated structure of this core
in the $N-S$ direction (Fig.~\ref{Fg3})
with an FWHP angular size of 
$\sim 45''\times57''$ (deconvolved), or $\sim 0.04\times0.06$ pc$^2$ linear size at $D = 203$ pc. 
The coordinates of the \nhhh\ peak,
the peak temperature \Tmb, its radial velocity $V_{\scriptscriptstyle \rm LSR}$, and the linewidth
$\Delta v$ (FWHM) are listed in Table~\ref{tbl-1}. The \nhhh\ spectrum is shown in Fig.~\ref{afg3}.
We did not detect any emission at the expected position in the (2,2) transition and, therefore,
only an upper limit on the rotational temperature is given in Table~\ref{tbl-3}. 
The $1\sigma$ upper limit on \Trot\ is very low, $< 10$~K. 
A noticeable feature
of the \object{Do279P8} spectrum is a very narrow linewidth of the hfs 
transitions, $\Delta v \approx 0.26$ \kms. 
Another remarkable characteristic is slow rotation of the cloud illustrated in Fig.~\ref{Fg3}{\bf b}:  
at a radius of $\sim 20''$ off the center ($\approx 0.02$ pc), the tangential velocity
does not exceed $\pm0.04$ \kms, which leads to an angular velocity
$\dot{\phi} \sim 6\times10^{-14}$ s$^{-1}$. 
The axis of the velocity centroid is slightly
tilted with respect to the main $N-S$ axis of the \nhhh(1,1) map.

The measured excitation temperatures at the offsets $(0'',0'')$ and $(0'',-40'')$
are close to the upper limits on the rotational temperatures (see Table~\ref{tbl-3}).
This may imply that the (1,1) transition is almost thermalized 
and ammonia emission
arises from an extended structure which fills the telescope beam homogeneously, $\eta \approx 1$.
At our angular resolution of $40''$, the \nhhh(1,1) hfs components do not show any
blueward or redward asymmetry indicating contracting or expanding motions, respectively.
Most probably, the core \object{Do279P8} is in an early stage of evolution.
According to Lee \& Meyers (2011), the starless cores evolve from the static to
the expanding and/or oscillating stages, and finally  to the contracting cores.
The suggestion that the core is in an initial stage of evolution is also supported
by low values of $\beta \la 0.1$ and $\beta' \approx 0.01$. 
In this estimation, we use 
$n_{\rm gas} \ga 1\times10^4$ \cmm\ ([\nhhh]/[H$_2$] $\sim 5\times10^{-7}$)
as a lower bound on the gas density.
With the observed diameter $d \sim 50''$ (deconvolved) and assuming spherical geometry, 
the lower limit on the mass of \object{Do279P8} is $M \sim 0.06M_\odot$.

\subsection{\object{Dobashi~279~P12} (\object{Do279~P12})}
\label{sect-4-4}

The source \object{Dobashi~279~P12}  (\object{Do279P12}, herein)
is a dark cloud with the opacity class
4-5 in accord with Lynds (1962). The cloud is known under the names 
\object{L572} (Lynds 1962),
\object{BDN~31.57+5.37} (Bernes 1977), or
the Serpens dark cloud which is a site of active star formation (Strom \etal\ 1974).
A large number of articles  has been devoted to studying this cloud in all spectral ranges
from radio to X-ray (for a review, see, e.g., Eiroa \etal\ 2008).

Single-dish ammonia observations of the Serpens dark cloud were started by Little \etal\ (1980)
and by Ho \& Barrett (1980).
Later on, Ungerechts \& G\"usten (1984) mapped an extended ammonia structure of this cloud with
the Effelsberg 100-m telescope.
Ammonia interferometric observations with the VLA by Torrelles \etal\ (1992) 
were directed to the study of the embedded high-density
molecular gas around the triple radio source Serpens \object{FIRS 1}~--- a typical Class~0 protostar
(Snell \& Bally 1986; Rodr\'igues \etal\ 1980).
Further interferometric (VLA) and additional single dish (Haystack 37-m, Effelsberg 100-m)
observations of \object{FIRS 1} by Curiel \etal\ (1996) 
revealed high-velocity ammonia 
emission (up to $\sim 30-40$ \kms\ from the line center) aligned with the
radio continuum jet. 

\begin{figure}
\vspace{0.0cm}
\hspace{-1.0cm}\psfig{figure=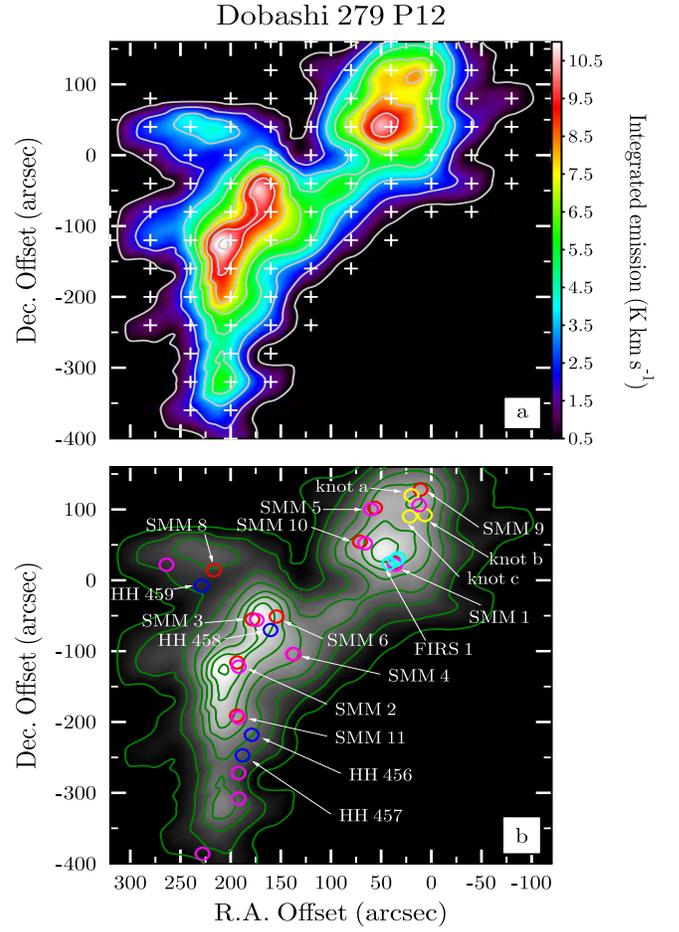,height=14.0cm,width=11.0cm}
\vspace{-0.7cm}
\caption[]{
({\bf a}): Same as Fig.~\ref{Fg1}{\bf a} but for the source \object{Do279P12}. 
The contour levels start at 0.5 K~\kms\ and increments are 1.5 K~\kms. 
({\bf b}): \nhhh(1,1) intensity map as in panel {\bf a} and positions of IR sources.
Open circles mark the Herbig-Haro objects (blue),
far-infrared continuum sources (red and yellow) (Davis \etal\ 1999),
submm-continuum sources (magenta) detected by SCUBA (Di Francesco \etal\ 2008), 
and the triple radio source \object{FIRS 1} (cyan) (Snell \& Bally 1986).
}
\label{Fg4}
\end{figure}

\begin{figure}
\vspace{-0.7cm}
\hspace{0.0cm}\psfig{figure=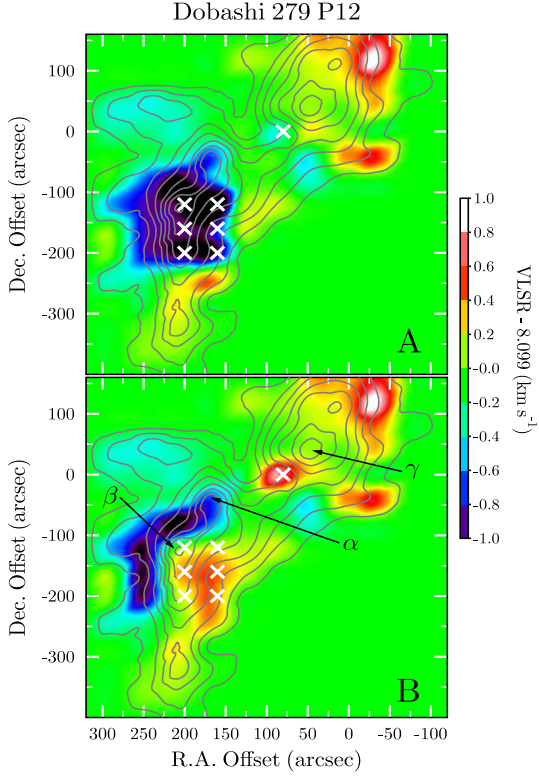,height=12.0cm,width=9.0cm}
\vspace{-1.0cm}
\caption[]{
\nhhh(1,1) intensity (grey contours) and radial velocity (color map) 
structure in the source \object{Do279P12}.
The velocity field is shown after subtraction of the mean radial
velocity $\langle V_{\scriptscriptstyle\rm LSR} \rangle = 8.099$ \kms.
The contour levels are the same as in Fig.~\ref{Fg4}{\bf a}.
Panels A and B represent the velocity maps for the components A and B in the 
\nhhh\ profile, respectively (see Fig.~\ref{afg4}).  
The white crosses mark offsets where the splitting of the \nhhh\ profiles is
detected. 
The mean radial velocities of the components A and B are
$\langle V_{\rm A} \rangle = 7.27\pm0.11$ \kms\ and 
$\langle V_{\rm B} \rangle = 8.54\pm0.08$ \kms. 
In panel B, the intensity peaks are labeled in accord with Table~\ref{tbl-1}. 
}
\label{Fg5}
\end{figure}

\begin{figure}
\vspace{0.0cm}
\hspace{0.5cm}\psfig{figure=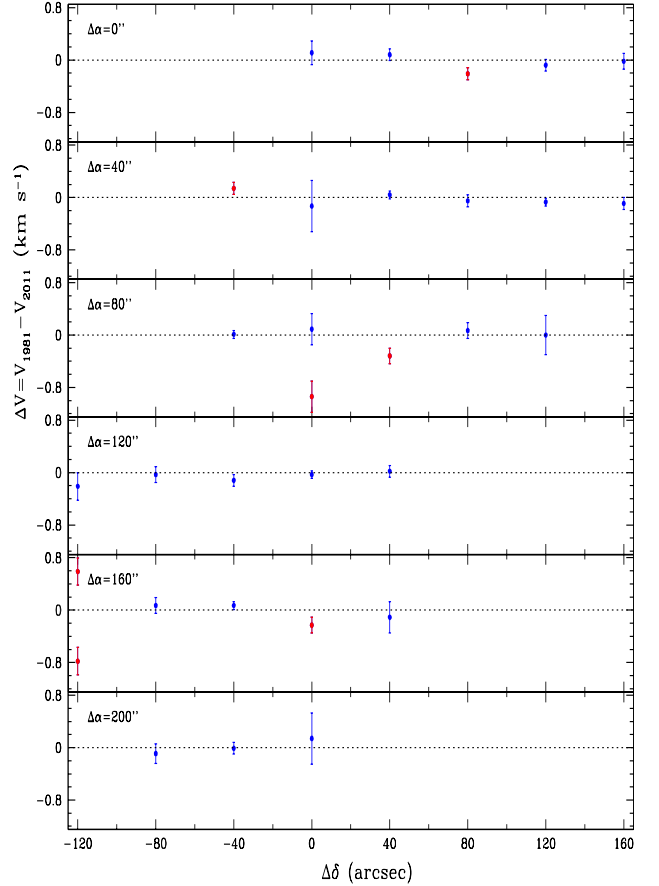,height=12.0cm,width=8.5cm}
\vspace{0.0cm}
\caption[]{
Dots with $3\sigma$ error bars mark the
differences between the radial velocities measured 
with the Effelsberg 100-m telescope in 1977-1981 
(Ungerechts \& G\"usten 1984) and in the present dataset
toward the same positions in \object{Do279P12}.
The red points indicate deviations from zero which are larger than $3\sigma$. 
At two offsets, $(\Delta\alpha,\Delta\delta) = (80'',0'')$ and $(160'',-120'')$, we detected double \nhhh\
profiles (see Fig.~\ref{afg4}). 
}
\label{Fg6}
\end{figure}

The results of our Effelsberg observations are shown in Figs.~\ref{Fg4}-\ref{Fg6}.
The \nhhh(1,1) map (Fig.~\ref{Fg4}) exhibits a complex structure with three bright peaks,
$T_{\scriptscriptstyle\rm MB} \sim 3$~K, located 
NW--SE at the offsets $\alpha = (160'',-40'')$, $\beta = (200'',-120'')$, and 
$\gamma = (40'',40'')$ (see Table~\ref{tbl-1} and Fig.~\ref{Fg5}, panel B). 
The $\alpha$ peak lies very close to the maximum of the \nhhh\ emission detected by Little \etal\ (1980),
whereas the $\beta$ and $\gamma$ peaks are spatially coincident with positions of the two peaks
localized by Ho \& Barrett (1980).
The measured \nhhh\ profiles show in this case a double structure in both the
(1,1) and (2,2) transitions at four positions (Fig.~\ref{afg4}, Table~\ref{tbl-5}) 
and a split (1,1) line at another three positions where the (2,2) line 
was not detected. 
This multicomponent structure has not been resolved previously in the 
ammonia observations cited above.
Since the splitting is also seen 
in the optically thin ammonia lines, the observed asymmetry of the \nhhh\ profiles
has clearly a kinematic origin.

We used a two-component model, Eq.~(\ref{Eq8}), to fit the data. The result is illustrated
in Fig.~\ref{afg4} by the red lines, the detected velocity subcomponents are labeled by
letters $A$ and $B$. The seven offsets with the split  \nhhh\ lines 
are marked by white crosses on the velocity maps shown in Fig.~\ref{Fg5}. 
The cluster of the white crosses at the SE part of \object{Do279P12} outlines
a small area of the bipolar outflow with an angular extent of $\sim 40'' \times 80''$. 
Table~\ref{tbl-5} displays the physical parameters estimated at different offsets. 
The mean radial velocities of the components A and B 
averaged over the seven offsets are
$\langle V_{\rm A} \rangle = 7.27\pm0.11$ \kms\ and 
$\langle V_{\rm B} \rangle = 8.54\pm0.08$ \kms, 
and the velocity difference is
$\Delta V_{\scriptscriptstyle\rm B-A} = 1.3\pm0.1$ \kms.
Panel $A$ in Fig.~\ref{Fg5} shows that the gas surrounding the SE double component region
has predominantly a negative radial velocity with respect to $\langle V_{\scriptscriptstyle\rm LSR} \rangle$,
whereas the gas shown in panel $B$ has a positive $V_{\scriptscriptstyle\rm LSR}$. 
The observed velocities converge to the mean $\langle V_{\scriptscriptstyle\rm LSR} \rangle$ at the
cloud's outskirts. 

Figure~\ref{Fg5} shows also a velocity gradient along the E--W direction
around the isolated double component region at $(80'',0'')$ with $V_{\rm A} = 7.89$ \kms\
and $V_{\rm B} = 8.92$ \kms.
The adjoint regions at $(120'',0'')$ and $(40'',0'')$ have 
intermediate velocities of 
$V_{\scriptscriptstyle\rm LSR} = 8.14$ and 8.42 \kms, respectively,
which become closer to the mean radial velocity of the cloud 
$\langle V_{\scriptscriptstyle\rm LSR} \rangle = 8.099$ \kms\ at larger offsets:
$V_{\scriptscriptstyle\rm LSR} = 7.90$ and 8.04 \kms\ at
$(280'',0'')$ and $(-40'',0'')$, respectively.

The measurement of the excitation temperatures reveals \Tex\ essentially lower than \Trot\ at all
23 offsets listed in Table~\ref{tbl-5}. As discussed above, this may indicate clumpiness of the
ammonia distribution and $\eta < 1$. 
In Fig.~\ref{Fg4}{\bf b}, we also show 
the Herbig-Haro (HH) objects and far-infrared continuum sources taken from Davis \etal\ (1999),
the submm-continuum sources detected by SCUBA (Di Francesco \etal\ 2008), and the position
of the triple radio source \object{FIRS 1} (Snell \& Bally 1986) which coincides with 
the brightest infrared object SMM1.
The closest position of the split \nhhh\ lines to the bipolar outflow in \object{FIRS 1} is
$(80'',0'')$ which corresponds to a projected 
angular distance of $\theta = 45''$ from the SE knot of the radio source \object{FIRS 1} 
(see also Fig.~13 in Snell \& Bally 1986).  
All other splittings are located in the vicinity of the $\beta$ peak,
around the SMM2 and SMM11 sources. 

The observed linewidths \Dv\ of the split \nhhh\ components
are in the range between  0.62 \kms\ and 0.90 \kms. 
However, larger linewidths ($\Delta v \ga 1.0$ \kms) 
measured at different offsets (Table~\ref{tbl-5})
may imply unresolved subcomponents of the ammonia lines due to the moderate spectral
resolution of our database. Gas flows and jets usually accompany the mass accretion processes in YSOs and
protostars. From this point of view the split and/or widened line profiles of \nhhh,~--- a tracer of the
dense cores,~--- may indicate the first stages of the accretion processes in protostars,
where jets of HH objects have not yet been formed. 

It may be illustrative to compare our $V_{\scriptscriptstyle\rm LSR}$
values with those measured with the same telescope and beam size in 1977-1981 (Ungerechts \& G\"usten 1984). 
In total, we have 29 overlapping
positions shown in Fig.~\ref{Fg6}. 
The calculated differences $\Delta V = V_{1981} - V_{2011}$ are marked by
dots with $3\sigma$ error bars. The red points indicate deviations from zero which are larger than $3\sigma$. At two
positions, $(\Delta\alpha,\Delta\delta) = (80'',0'')$ and $(160'',-120'')$, we detected double \nhhh\
profiles. Among these 29 positions, significant deviations from zero are revealed at 6 offsets, but other points show
no variations over this time scale. The ammonia line splitting was not seen at $(80'',0'')$ in 1981, although the
linewidth was $\Delta v_{1981} = 0.85 \pm 0.20$ \kms, 
which is comparable to that given in Table~\ref{tbl-5}. 
However, at $(160'',-120'')$ the linewidth was $\Delta v_{1981} = 1.68 \pm 0.17$ \kms, 
i.e., two times larger than the value from Table~\ref{tbl-5}.
We note that the weighted mean of the linewidth at the overlapping positions is 
$\langle \Delta v \rangle_{1981} = 0.93$ \kms,
and the dispersion of the sample is $\sigma_{\Delta v} = 0.18$ \kms.
The enhanced \Dv(\nhhh) at 
$(160'',-120'')$ observed 30 years ago was probably caused by the unresolved structure
of the ammonia lines. The maximum deviation 
$\Delta V_{\scriptscriptstyle \rm LSR} \sim 1$ \kms\ (Fig.~\ref{Fg6}) constrains the
gas acceleration in an outflow at the level of $\dot{V} \la 0.03$ km~s$^{-1}$~yr$^{-1}$.
This means that the high-velocity ammonia emission up to $\sim 30-40$ \kms\ from
the line center, observed, for example, in 
\object{FIRS 1} (Curiel \etal\ 1996), could be reached in a period of
$\sim 1000$ yr. This is comparable to the typical time scale of $10^3$ yr~--- a scale for
the episodic ejection of material from protostars (e.g., Ioannidis \& Froebrich 2012).
The mass of the protostar should be $\la 0.3M_\odot$ in order to eject the gas clump
at the 40 AU radius with the acceleration $\dot{V} \sim 0.03$ km~s$^{-1}$~yr$^{-1}$.

\subsection{\object{Dobashi~321~P2} (\object{Do321~P2})}
\label{sect-4-5}

The source \object{Dobashi~321~P2} (\object{Do321P2}, herein) 
is a filamentary dark cloud with the opacity class 6
in accord with Lynds (1962). The cloud is known under the names 
\object{L673} (Lynds 1962), \object{BDN~46.22-1.34} (Bernes 1977), or \object{P77} (Parker 1988).
\object{Do321P2}  was observed in ammonia lines by 
Anglada \etal\ (1997) and Tobin \etal\ (2011). 

\begin{figure}
\vspace{0.0cm}
\hspace{-3.5cm}\psfig{figure=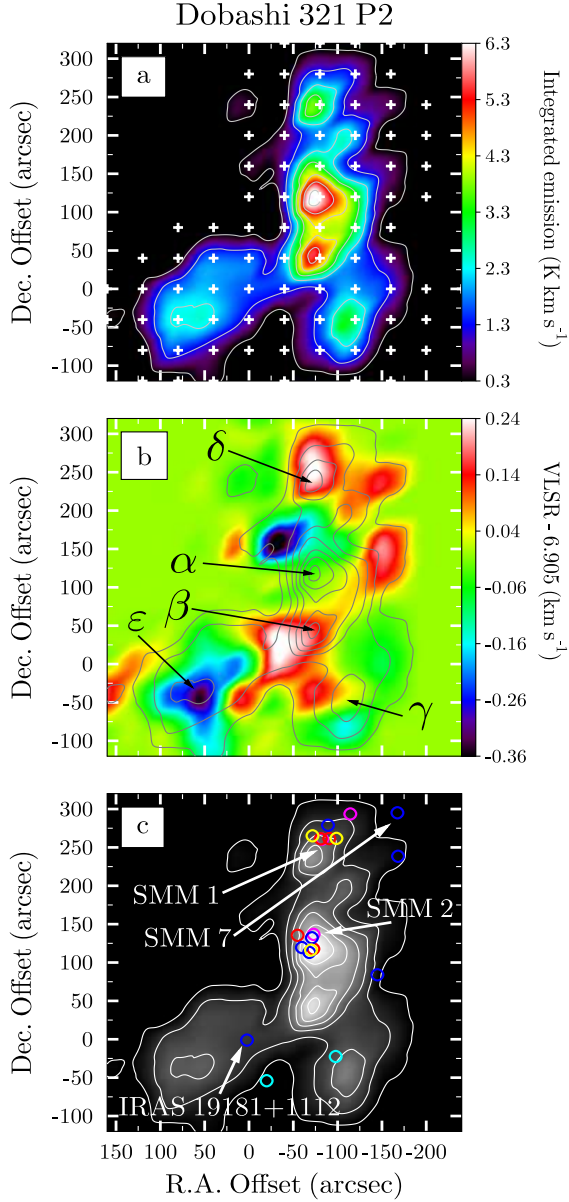,height=20.0cm,width=18.0cm}
\vspace{-3.0cm}
\caption[]{
({\bf a}): Same as Fig.~\ref{Fg1}{\bf a} but for the source \object{Do321P2}.
The contour levels start at 0.3 K~\kms\ and increments are 1.0 K~\kms.
({\bf b}): \nhhh(1,1) radial velocity (color map) structure. 
The velocity field is shown after subtraction of the mean radial
velocity $\langle V_{\scriptscriptstyle\rm LSR} \rangle = 6.905$ \kms.
The intensity peaks are labeled in accord with Table~\ref{tbl-1}. 
({\bf c}): \nhhh(1,1) intensity map  as in panel {\bf a}. 
Open circles of different colors mark 
Class~I (red), intermediate type (yellow), Class~II (blue), and Class~III (cyan) YSOs
taken from Tsitali \etal\ (2010). The SCUBA sources from Visser \etal\ (2002)
are marked by magenta: SMM7 is a starless core, whereas SMM1 and SMM2 are
protostars which coincide with \object{IRAS 19180+1116} and \object{IRAS 19180+1114},
respectively.
}
\label{Fg7}
\end{figure}

Our \nhhh\ observations are shown in Fig.~\ref{Fg7}. 
The ammonia structure consists of five main subcondensations. Their 
reference positions are listed in Table~\ref{tbl-1}.
The brightest $\alpha$ and $\beta$ peaks with $T_{\scriptscriptstyle\rm MB} \sim 3$~K 
are located near the sources \object{IRAS 19180+1114} (SMM2) and
\object{IRAS 19180+1116} (SMM1), respectively. 
The ammonia peak detected by Anglada \etal\ has an offset $(-123'',83'')$
and is resolved into two clumps $\alpha$ and $\beta$ in our observations (see Fig.~\ref{Fg7}).
We also detected three weaker ammonia
clumps $\gamma$, $\delta$, and $\varepsilon$ with $T_{\scriptscriptstyle\rm MB} \sim 1.7,
1.6$, and 1.3~K, respectively.
The $\gamma$ and $\varepsilon$ condensations, 
localized within the confines of the older map of Anglada \etal\ (1997),
peak near the source \object{IRAS 19181+1112}, whereas the $\delta$ clump was not 
resolved by Anglada \etal\ (1977) because of a too coarse angular resolution. 

Interferometric observations of a small region around SMM2
($\sim 100'' \times 100''$) in lines of N$_2$H$^+$ and \nhhh\ (Tobin \etal\ 2011; see Fig.~14)
show a substantial amount of extended ammonia emission 
which closely traces the spacial distribution of N$_2$H$^+$.
The measured \nhhh\ radial velocity \VLSR\ = 6.93 \kms, the linewidth $\Delta v = 0.50$ \kms,
and the column density $N$(\nhhh) = $7.2\times10^{14}$ \cm\ (Table~8 in Tobin \etal)
are in a good agreement with our values for the $\alpha$ peak (Table~\ref{tbl-1}):
\VLSR\ = 6.88 \kms, $\Delta v = 0.49$ \kms, and $N$(\nhhh) = $6 \times10^{14}$ \cm. 
However, the estimates of the excitation temperature
at this position contradict sharply: 
\Tex\ = 33.3~K (Tobin \etal) and 6.8~K (our value from Table~\ref{tbl-3}).
We detected \nhhh(1,1) and (2,2) at the $\alpha$ and $\beta$ peaks (Fig.~\ref{afg5}).
Their profiles are well described by a single-component model (red lines in Fig.~\ref{afg5}, model parameters
are given in Table~\ref{tbl-3}). 
Both spectra show the same \Tex, which is also consistent with the values measured in other
dense molecular cores (see Tables~\ref{tbl-4}-\ref{tbl-6}).  

The \nhhh\ velocity map (Fig.~\ref{Fg7}{\bf b}) shows irregular structure with 
distinct pockets of redshifted and blueshifted ammonia 
emission around the clusters of YSOs at the positions of SMM1 and SMM2 (Fig.~\ref{Fg7}{\bf c}).
The observed kinematics are caused by large-scale motions in this star-forming region which cannot
be interpreted unambiguously.

The apparent diameters (beam deconvolved) 
of the two \nhhh\ condensations $d_\alpha = 60''$ and $d_\beta = 40''$ and the
gas densities listed in Table~\ref{tbl-3} allow us to estimate masses of these clumps assuming
spherical geometry: $M_\alpha \sim 0.3 M_\odot$, and $M_\beta \sim 0.1 M_\odot$. The relative
ammonia abundance in both clumps is [\nhhh]/[H$_2$] $\sim 2\times10^{-7}$.
We note that the corresponding virial masses of these clumps are considerably larger 
($M^\alpha_{\rm vir} \sim 2M_\odot$, and $M^\beta_{\rm vir} \sim 1M_\odot$) and probably affected
by the masses of YSOs embedded in the cloud.

\subsection{\object{Kawamura~05} (\object{Ka05})}
\label{sect-4-6}

The source \object{Kawamura~05} (\object{Ka05}, herein) 
is a Bok globule of the opacity class 6 (Lynds 1962). 
The cloud is known under the names 
\object{B335} (Barnard 1927), \object{L663} (Lynds 1962), or \object{CB199} (Clemens \& Barvainis 1988).
Being the prototype of an isolated, star-forming dark cloud,
\object{B335} has been studied at many wavelengths.

Single-pointing ammonia observations towards \object{B335} 
were carried out by Ho \etal\ (1977, 1978), Myers \& Benson (1983), 
Benson \& Myers (1989), and Levshakov \etal\ (2010). 
The first mapping of this core in the \nhhh(1,1) and (2,2) lines with the Effelsberg
telescope revealed an elongated N-S structure (Menten \etal\ 1984, M84 herein). 
The kinetic temperature measured at four positions, where the \nhhh(2,2) line
was observed, is uniformly distributed over the core region,
\Tkin\ = 10-12~K.  

\begin{figure}
\vspace{0.0cm}
\hspace{-1.0cm}\psfig{figure=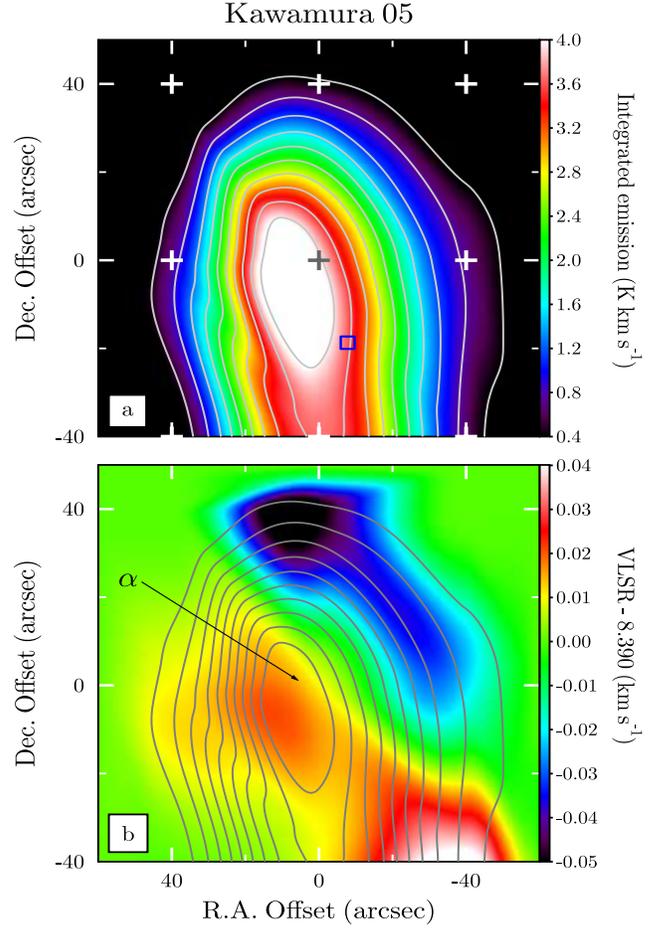,height=14cm,width=11cm}
\vspace{-0.7cm}
\caption[]{
({\bf a}): Same as Fig.~\ref{Fg1}{\bf a} but for the source \object{Ka05}. 
The contour levels start at 0.4 K~\kms\ and increments are 1.0 K~\kms.  
The square indicates the position of the far-infrared source IRAS 19345+0727
coinciding with the 3.6 cm continuum source (Anglada \etal\ 1992).
({\bf b}): \nhhh(1,1) radial velocity (color map) structure. 
The velocity field is shown after subtraction of the mean radial
velocity $\langle V_{\scriptscriptstyle\rm LSR} \rangle = 8.390$ \kms.
The intensity peak is labeled in accord with Table~\ref{tbl-1}. 
}
\label{Fg8}
\end{figure}

We mapped the core \object{Ka05} at nine offsets around the reference point
$\alpha$(J2000) = 19:37:01.3, $\delta$(J2000) = +07:34:29.6 which is shifted 
off the center position in M84 at 
$(\Delta \alpha, \Delta \delta) = (-24.0'', -42.8'')$.
The results of our observations are shown in Fig.~\ref{Fg8}.
Two positions $(0'', 0'')$ and $(0'', -40'')$ with 
\Tmb\ = 2.5~K and 2.1~K (Table~\ref{tbl-1}) 
coincide, respectively, with the offsets $(20'', 40'')$ with \Tmb\ = 3.4~K and 
$(20'', 0'')$ with \Tmb\ = 2.8~K in M84. In between these positions,
at $(20'', 20'')$, the brightness temperature was even higher, \Tmb\ = 4.1~K (M84),
but we did not observe this point with our larger step size of $40''$,
as well as the point $(20'', 60'')$ where \Tmb\ = 3.1~K.
However, we measured \nhhh(1,1) at $(0'', 40'')$ with \Tmb\ = 0.6~K which
corresponds to the point $(20'', 80'')$ not listed in Table~1 in M84. The
opposite marginal point $(20'', -20'')$ along the cut $\Delta \alpha = 20''$ in M84
has \Tmb\ = 0.8~K. This indicates sharp drops of the brightness temperature at the
edges of the disk-like envelope surrounding the protostar.

The spectra with the measured (1,1) and (2,2) lines are shown in Fig.~\ref{afg6}.
The radial velocities at two offsets $(0'', 0'')$ and $(0'', -40'')$  
are the same as they were
in 1982/83 (cf. Fig.~\ref{Fg6}, Sect.~\ref{sect-4-4}, where velocity shifts at some positions
in the map of \object{Do279P12} were detected for the same three decades).
The linewidths (Table~\ref{tbl-3}) are also consistent with M84. 
At these offsets, the measured kinetic temperatures,
excitation temperatures, total optical depths, and column densities
are in line with M84.  

The \nhhh(1,1) velocity map is shown in Fig.~\ref{Fg8}{\bf b}.
The velocity distribution resembles a rigid body rotation 
with the tangential velocity increasing from $\approx -0.040$ \kms\ (NE) 
to $\approx +0.040$ \kms\ (SW) at cloud flanks. 
The radius of the \nhhh\ disk-like envelope is about $40''$ in accord with our and the M84 maps.
The distance to \object{B335} was recently redetermined to be only
$D = 90-120$ pc (Olofsson \& Olofsson 2009).
This gives us a radius $r \sim 0.02$~pc,
and an angular velocity  $\dot{\phi} \sim (6-8)\times10^{-15}$ s$^{-1}$ ($\beta \la 0.001$), 
which is close to the value of M84, $\dot{\phi} \sim 1.5\times10^{-14}$ s$^{-1}$.
The corresponding stability parameter $\beta' \approx 2\times10^{-4}$. 
With the apparent diameter $d \sim 0.04$~pc, the relative abundance ratio is 
[\nhhh]/[H$_2$] $\sim 4\times10^{-7}$.

To compare with previous estimates of the mass of the gas giving rise to ammonia
emission, we assume spherical geometry and for 
the gas density $n_{{\rm H}_2} \sim 2\times10^4$ \cmm\ (Table~\ref{tbl-3}) 
to obtain $M \sim 0.05M_\odot$~--- 40 times lower than
the value reported by Benson \& Myers (1989) who used a core radius $R = 0.11$ pc ($D = 250$ pc)
and a gas density $n_{{\rm H}_2} \sim 6\times10^3$ \cmm.

If the centrifugal forces prevent the disk-like structure from collapse,
then the mass of the center source is given, within a factor of 2,  
by (e.g., Frerking \& Langer 1982):
\begin{equation}
M_c = V^2_t R /G,
\label{Eq22}
\end{equation}
where $R$ is the distance of the material from the source and $G$ is the gravitational constant.
For $V_t = 0.040$ \kms, it gives $M_c \sim 0.03M_\odot$.
This estimate indicates that \object{Ka05} is a low-mass protostellar source which agrees
with the measurement of the central source mass $M_c \sim 0.04M_\odot$ from SMA observations of
the C$^{18}$O(2-1) emission (Yen \etal\ 2010).

\subsection{\object{Dobashi~279~P7} (\object{Do279~P7})}
\label{sect-4-7}

\begin{figure}
\vspace{0.0cm}
\hspace{-1.0cm}\psfig{figure=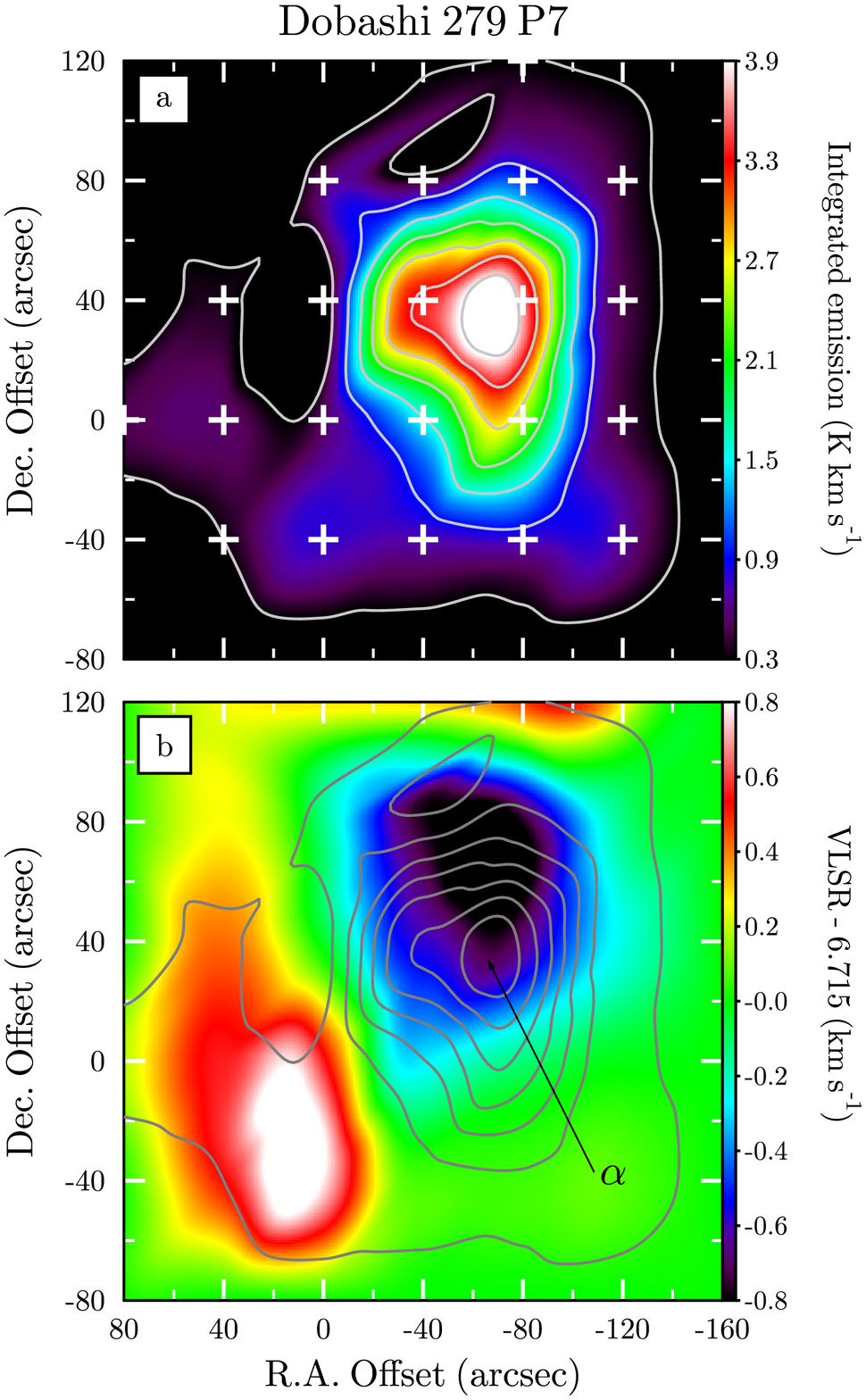,height=14cm,width=11cm}
\vspace{-0.7cm}
\caption[]{
({\bf a}): Same as Fig.~\ref{Fg1}{\bf a} but for the source \object{Do279P7}. 
The contour levels start at 0.3 K~\kms\ and increments are 0.6 K~\kms. 
({\bf b}): \nhhh(1,1) radial velocity (color map) structure. 
The velocity field is shown after subtraction of the mean radial
velocity $\langle V_{\scriptscriptstyle\rm LSR} \rangle = 6.715$ \kms.
The intensity peak is labeled in accord with Table~\ref{tbl-1}. 
}
\label{Fg9}
\end{figure}

The dense molecular cloud \object{Dobashi~279~P7} (\object{Do279P7}, herein)
is the next example of a starless core from our list. 
This is a newly discovered dense clump which is so far poorly studied.
No embedded IRAS sources or pre-main-sequence stars are known within $80''$ 
off the core center.
The \nhhh(1,1) intensity distribution shown in Fig.~\ref{Fg9} has 
approximately circular symmetry.
The half-power dimension of the core is $80'' \times 80''$, or, correcting for our beam,
$70'' \times 70''$, i.e., the source is resolved in both right ascension and declination.
Assuming a distance to the source of $D \approx 203$ pc (see Sect.~\ref{sect-1}), 
its radius is about 0.03 pc. 
The parameters for the density peak at $(-80'',40'')$ are given in Table~\ref{tbl-1}. 
The \nhhh\ spectra are shown in Fig.~\ref{afg7}.
We detected the (2,2) transition at four positions which allow us to estimate
\Trot\ $= 11.5-13.5$~K and \Tkin\ = $12.4-14.9$ (Table~\ref{tbl-3}). 
These values exceed slightly the temperatures measured in the starless core 
\object{Ka01}, but the linewidths in \object{Do279P7},
are significantly larger than the linewidths 
in \object{Ka01} and \object{Do279P8}~-- another starless core
but without detected (2,2) transition (see Table~\ref{tbl-3}).
The wider lines in \object{Do279P7} indicate a higher level of turbulence in this core
with the ratio of turbulent to thermal velocity dispersion of  
$\Delta v_{\rm turb}/\Delta v_{\rm th} \approx 5$. 
The profiles of the \nhhh(1,1) hfs components do not show any
blueward or redward asymmetry indicating contracting or expanding motions.
However, the \nhhh\ intensity map shows a filament-like
structure in the S-E direction (P.A. $\approx 135^\circ$) with \VLSR\ $\sim 0.8$ \kms\
(Fig.~\ref{Fg9}). 

The core exhibits a linear velocity gradient from 
the southern (\VLSR\ $\sim 0.0$ \kms) to the northern edge (\VLSR\ $\sim -0.8$ \kms)
that can be interpreted 
as rigid body rotation around the axis $\Delta \delta \approx 20''$, P.A. $\approx 90^\circ$
(Fig.~\ref{Fg9}{\bf b}).
The measured tangential velocity of $\pm0.4$ \kms\ at $r \approx 0.03$~pc 
leads to an angular 
velocity of $\dot{\phi} \approx 5\times10^{-13}$ s$^{-1}$. 
With $n_{{\rm H}_2} \sim 1\times10^4$ \cmm\ (Table~\ref{tbl-3}) and assuming spherical geometry, one obtains
a core mass of $M \sim 0.2M_\odot$. 
The estimates of the angular velocity and the gas density give $\beta \sim 6$ 
which shows that the rotation energy is larger or comparable to the gravitational energy of this starless core.
The parameter $\beta' \approx 0.3$ is also rather high as compared with other isolated cores 
\object{Ka01}, \object{Do279P8}, and \object{Ka05} where $\beta' \ll 1$. 
The measured column density $N \sim 6\times10^{14}$ \cm\ and the
linear size $d \sim 0.06$~pc indicate a relative ammonia abundance of
[\nhhh]/[H$_2$] $\sim 3\times10^{-7}$.

\subsection{ \object{Serpens South 3} (\object{SS3}) }
\label{sect-4-8}

The source \object{Serpens South 3} (\object{SS3}, herein) 
is a dense core located in an active star-forming region
discovered by Gutermuth \etal\ (2008) in {\it Spitzer} IRAC mid-infrared imaging of the Serpens-Aquila rift.
The central part of this region,  near $\alpha$~= 18$^{\rm h}$\ 30$^{\rm m}$\ 03$^{\rm s}$, 
$\delta$~= $-02^\circ$\ 01$'$\ 58.2$''$ (J2000),  
which is called Serpens South, is composed of 37 Class~I and 11 Class~II sources
at high mean surface density ($> 430$ pc$^{-2}$) 
and short median nearest neighbor spacing (0.02 pc).
The overall structure of the star-forming region exhibits a typical filamentary dense
gas distribution elongated for about $14'$ in the NW-SE direction.

\begin{figure}
\vspace{0.0cm}
\hspace{-3.5cm}\psfig{figure=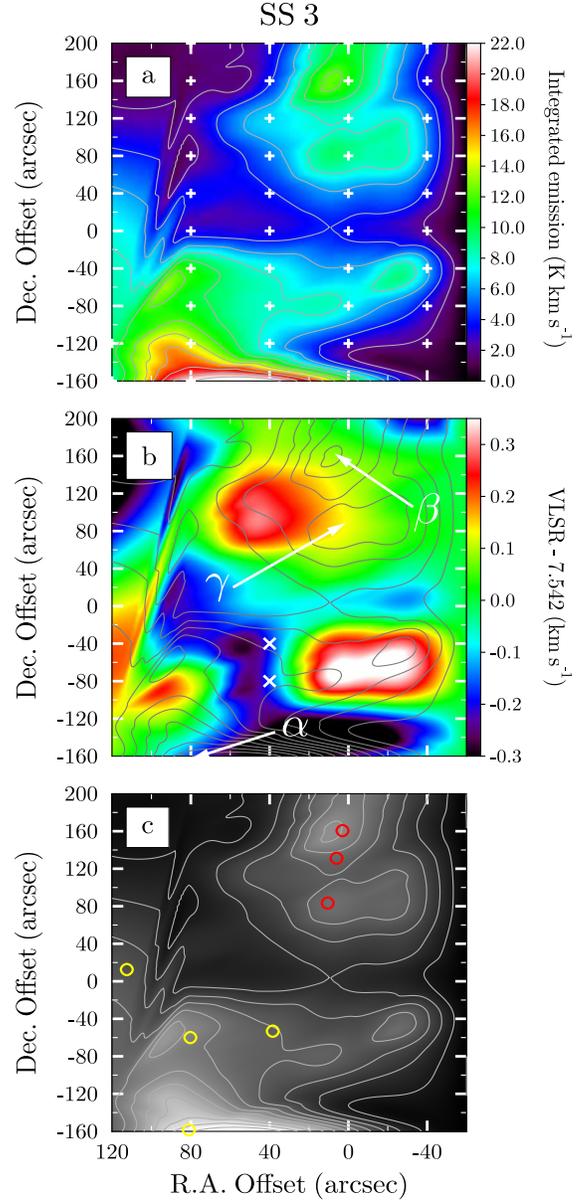,height=20cm,width=18cm}
\vspace{-3.0cm}
\caption[]{
({\bf a}): Same as Fig.~\ref{Fg1}{\bf a} but for the source \object{SS3}. 
The contour levels start at 0.0 K~\kms\ and increments are 2.0 K~\kms. 
({\bf b}): \nhhh(1,1) radial velocity (color map) structure. 
The velocity field is shown after subtraction of the mean radial
velocity $\langle V_{\scriptscriptstyle\rm LSR} \rangle = 7.542$ \kms.
The white crosses mark offsets where the splitting of the \nhhh\ profiles is detected 
(see Fig.~\ref{afg8}).
The mean radial velocities of the components A and B are
$\langle V_{\rm A} \rangle = 7.865\pm0.005$ \kms\ and 
$\langle V_{\rm B} \rangle = 6.856\pm0.010$ \kms. 
The intensity peaks are labeled in accord with Table~\ref{tbl-1}. 
({\bf c}): \nhhh\ intensity map as in panel {\bf a}
and positions of the MAMBO millimeter sources
taken from Table~1 in Maury \etal\ (2011).
Red and yellow open circles mark starless and YSO candidates,
respectively. 
}
\label{Fg10}
\end{figure}

For the first time, we mapped the core \object{SS3} in the \nhhh(1,1) and (2,2) lines
around the reference point 
$\alpha$ = 18$^{\rm h}$\ 29$^{\rm m}$\ 57.1$^{\rm s}$, 
$\delta$ = $-02^\circ$\ 00$'$\ 10.0$''$ (J2000)
in the area $\Delta\alpha \times \Delta \delta = 160'' \times 360''$ (Fig.~\ref{Fg10}).
The nominal center of the \nhhh\ map  is shifted 
by $2'29''$ (P.A. $\approx -40^\circ$)
with respect to the center of Box~1 in the CO map of Nakamura \etal\ (2011, herein N11)
and coincides with the intensity peak of the CO outflow lobe marked by ${\it R2}$ in N11. 
The angular size of the CO lobe within the 15 K~\kms\ contour
is about $70'' \times 60''$ (Figs.~9 and 15 in N11), and the whole region
lies in the lowest \nhhh\ intensity `valley' running  in E-W direction
along the cut $\Delta \delta = 0''$ in Fig.~\ref{Fg10}.
To the south of this valley, at $\Delta \delta \approx -140''$,
an intensity peak of \nhhh\ ($\alpha$ peak in Table~\ref{tbl-1} and Fig.~\ref{Fg10}{\bf b}) 
occurs at the northern part of another CO outflow lobe marked ${\it R3}$ in N11.
These two redshifted lobes ${\it R2}$ and ${\it R3}$ appear to have
the blueshifted counterparts ${\it B1}$ and ${\it B4}$  originating
from a compact 3 mm source~--- a deeply embedded, extremely
young Class~0 protostar (N11).
It is interesting to note that the configuration of the \nhhh\ valley just follows
the lowest dust emission distribution on the 1.1 mm continuum image of this area
shown in Fig.~15(b) in N11. To the north and to the south of the valley
there are regions of relatively strong 1.1 mm emission which are correlated with
strong \nhhh\ emission. The three \nhhh\ peaks $\alpha, \beta$, and $\gamma$ 
listed in Table~\ref{tbl-1} and labeled in Fig.~\ref{Fg10}{\bf b} 
have very high brightness temperatures \Tmb\ = 5.0, 3.1, and 3.0 K.

Examples of the observed \nhhh(1,1) and (2,2) spectra are shown in Fig.~\ref{afg8}.
Here again, as was the case of the core \object{Do279P12}, we find split \nhhh\
profiles. The apparent asymmetry of the ammonia lines is observed in optically
thin hf components and, thus, cannot be due to radiative transfer effects (self-absorption).
The components $A$ and $B$, detected at two offsets (see Table~\ref{tbl-6}), 
have the radial velocities 
$\langle V_{\rm A} \rangle = 7.865\pm0.005$ \kms\ and 
$\langle V_{\rm B} \rangle = 6.856\pm0.010$ \kms. 
The map  of the velocity field is shown in Fig.~\ref{Fg10}{\bf b} where the positions of the
split \nhhh\ profiles are labeled by the white crosses. 
The closest Class~0 protostar to these positions is the MAMBO mm source MM10
(Maury \etal\ 2011) marked by the yellow circle in Fig.~\ref{Fg10}{\bf c} at the offset $(38'',-53'')$.
However, since the whole region is very complex, it is not clear which source
drives this presumably bipolar molecular outflow causing the observed splitting of the
\nhhh\ profiles.

The measured linewidths listed in Table~\ref{tbl-6} show that the values $\Delta v \ga 0.8$ \kms\ 
are symmetrically distributed around the positions of the split \nhhh\ profiles
in the E-W strip of about $120''$ width, cut by the vertical edges of our map. 
The strip of enhanced linewidths (see Fig.~\ref{afg12}) may be associated with
the E-W bipolar outflow, thus giving rise to the local redshifted and blueshifted patches 
of the velocity map shown in Fig.~\ref{Fg10}{\bf b}. 

It is seen from Table~\ref{tbl-6} that the values of the measured parameters fluctuate considerably
from one position to another. Namely, the excitation temperature ranges from 3.3~K to 8.4~K,
\Trot~--- from 8.6~K to 14.8~K, \Tkin~--- from 8.9~K to 16.8~K, the gas density
$n_{{\rm H}_2}$~--- from $0.3\times10^4$ \cmm\ to $4.2\times10^4$ \cmm, 
the total optical depth $\tau_{11}$~--- from 1.5 to 20.0, and the column density
$N$(\nhhh)~--- from $0.2\times10^{15}$ \cm\ to $2.8\times10^{15}$ \cm.
We note that the highest gas density of $n_{{\rm H}_2} \sim 4.2\times10^4$ \cmm\ was measured
at two positions where we observed the split \nhhh\ lines profiles.
The largest optical depths with $13.2 \la \tau_{11} \la 20.0$ are localized along the N--S cut
$\Delta \alpha = 0''$ between $\Delta \delta = 80''$ and $200''$ 
which passes through the $\gamma$ and $\beta$ peaks in the region where three starless core candidates
(red circles in Fig.~\ref{Fg10}{\bf c}) were detected by Maury \etal\ (2011).
The maximum (16.6~K and 16.8~K) and minimum (8.9 K) kinetic temperatures are 
observed along the N--S cut at $\Delta \alpha = 80''$. The minimum of the kinetic 
temperature correlates with the minimum of the brightness temperature, \Tmb = 0.5~K, 
whereas \Tmb = 1.2~K at \Tkin = 16.8~K (the foot of the $\alpha$ peak)
and \Tmb = 5.0~K at \Tkin = 16.6~K (the top of the $\alpha$ peak). 
This gradient of \Tkin\ ($8.9-16.8$~K) is probably due to the presence of two YSOs localized at the top and
foot of the $\alpha$ peak shown in Fig.~\ref{Fg10}{\bf b}.

\subsection{Dobashi 243~P2, 279~P18, 279~P13, and 321~P1}
\label{sect-4-9}

The sources \object{Do243P2}, \object{Do279P18}, \object{Do243P13}, and \object{Do321P1}
(Table~\ref{tbl-1}) 
were observed at their central positions and at $40''$ offsets toward the four cardinal directions.
Only a weak \nhhh(1,1) emission (\Tmb\ $\la 1$~K) is seen 
toward each of them. The corresponding spectra are depicted in Fig.~\ref{afg9}.
For \object{Do321P1}, two possible detections of the ammonia emission
are shown at the bottom panels, but the noise is high
and the identification of the \nhhh(1,1) line is not certain. 
The measured radial velocity of \VLSR\ = 13.5 \kms\ for \object{Do321P1}
is somewhat larger than that measured for the other sources from Table~\ref{tbl-1} for which
\VLSR\ ranges between 6.0 \kms\ and 9.6 \kms, however, 
the $^{12}$CO(1-0) 115.3 GHz, $^{13}$CO(1-0) 110.2 GHz, and C$^{18}$O(1-0) 109.8 GHz 
emission lines were also detected at \VLSR\ $\sim 13$ \kms\ with 
the Delingha 14-m telescope (Wang 2012). 

No submillimeter-continuum sources have been found within $2'$ off the core centers of 
\object{Do279P18}, \object{Do243P13}, and \object{Do321P1}, but 
a SCUBA point-like source \object{J180447.8-043208} (Di Francesco \etal\ 2008) lies in the field of
\object{Do243P2} at the angular distance $\theta = 41''$, P.A. = $-30^\circ$ from the
nominal center $(0'',0'')$.

\subsection{Sources without detected \nhhh\ emission}
\label{sect-4-10}

A list of 37 targets from our dataset with undetected \nhhh\ emission is presented in Table~\ref{tbl-2}.
Columns 3 and 7 give the noise level per channel width of 0.077 \kms\ on a \Tmb\ scale and
the mapped area, respectively. In the fields of some targets there were found
the following objects. 

\object{Do279P21}: a point-like infrared source \object{IRAS 18191-0310} (Cutri \etal\ 2003) at
$\theta = 80''$, P.A. = $-174^\circ$.

\object{Do279P17}: a point-like infrared source \object{IRAS 18351+0010} (Cutri \etal\ 2003) at
$\theta = 17''$, P.A. = $74^\circ$.

\object{Do296P3}: a point-like submillimeter-continuum source \object{J185121.1-041714} (Di Francesco \etal\ 2008) at
$\theta = 29''$, P.A. = $-173^\circ$.

\object{Ka15}: a galaxy \object{gJ194852.5-103056} (Skrutskie \etal\ 2006) at
$\theta = 64''$, P.A. = $99^\circ$.

\object{Ka17}: a point-like infrared \object{IRAS 19470-0630} (Porras \etal\ 2003) at
$\theta = 57''$, P.A. = $162^\circ$.

\section{Summary}
\label{sect-5}

We have used the Effelsberg 100-m telescope to observe the 
\nhhh(1,1) and (2,2) spectral lines in high density molecular cores.
The targets were preliminarily selected from the
CO and $^{13}$CO survey of the Aquila rift cloud complex carried out with the Delingha 14-m telescope.
We measured  a minimum grid of 5 positions with $40''$ spacing centered upon the 
$(0'',0'')$ position determined from the maximum intensity of the CO maps.
We mapped a larger area for sources where ammonia emission was detected to delineate
the spatial distribution of \nhhh.

In total, we have mapped the first 49 CO sources (from $\sim 150$ targets) 
and detected 12 \nhhh\ emitters what gives an overall
detection rate of $\sim 24$ per cent 
which is 3 times lower as compared to the Pipe Nebula which was observed with the GBT 100-m telescope 
(Rathborne \etal\ 2008) and which is not too far from the clouds studied here (see Sect.~\ref{sect-1}). 
The \nhhh\ sources in our sample represent diverse populations of molecular clouds from 
isolated and homogeneous starless cores with suppressed turbulence
to star formation regions with complex intrinsic motion and gas density fluctuations.

The starless cores \object{Ka01}, \object{Do279P8}, \object{Do279P7}, \object{Do279P18}, \object{Do243P13},
and a candidate~--- the clump $\varepsilon$ \object{Do279P6}~--- 
have radial velocities \VLSR\ ranging from 6.0 \kms\
to 9.6 \kms\ with the mean $\langle$\VLSR$\rangle$ = 8.1 \kms, and dispersion $\sigma = 1.2$ \kms.
The only starless core with \VLSR\ beyond the $3\sigma$ interval around the mean is
\object{Do321P1} for which \VLSR\ = 13.5 \kms. 
Three cores~--- \object{Ka01}, \object{Do279P8}, and the clump $\varepsilon$ \object{Do279P6}~--- demonstrate rather narrow
linewidths, $\Delta v \la 0.4$ \kms, whereas the others have $\Delta v \sim 0.5-1$ \kms.  
The masses of the starless cores do not exceed a solar mass and are 
typically $(0.1-0.5)M_\odot$, except for the cores $\varepsilon$ \object{Do279P6}, \object{Do279P8}, and \object{Ka05} 
for which $M < 0.1M_\odot$.   
The measured 
ammonia abundance, $X=$ [\nhhh]/[H$_2$], ranges between $1\times10^{-7}$ and $5\times10^{-7}$ with
the mean $\langle X \rangle = (2.7\pm0.6)\times10^{-7}$. 

For three cores~--- \object{Ka01}, \object{Do279P8}, and \object{Ka05}~--- we measured 
low angular velocities (parameters $\beta$ and $\beta' \ll 1$) 
and their rotation energies are less than the gravitational energy,
but \object{Do279P7}~--- a fast rotator~--- yields a rotational energy comparable 
to the gravitational energy
($\beta \approx 6$ and $\beta' \approx 0.3$).
It also has a high ratio of the turbulent to thermal velocity.

Two isolated star-forming cores, \object{Ka05} and \object{Do243P2}, were observed. 
The cores' radial velocities are \VLSR\ = 8.4 \kms\ and 7.0 \kms, respectively, and, thus, both of them belong
to the same group of cores with \VLSR\ $\sim 8$ \kms. 
For the former core, we estimate a sub-solar mass and thus confirm that \object{Ka05}
is a low-mass protostellar source. Both cores exhibit narrow linewidths, $\Delta v \sim 0.3$ \kms.

Most intense ammonia 
emission was observed towards four clouds with complex gas density and velocity structures
which harbor numerous YSOs: 
\object{Do279P6}, \object{Do279P12}, \object{Do321P2}, and \object{SS3}.
Each of them consists of a number of clumps with different angular sizes. 
The $\langle$\VLSR$\rangle$ values show that the complex clouds have the same radial velocities as the
starless cores and isolated star-forming clouds from our dataset.
It is only one cloud~---\object{Do279P12}~--- where the dispersion of bulk motions is comparable to the 
ammonia linewidths. The internal bulk motions in the other three clouds are less pronounced and
show lower dispersions as compared to the \nhhh\ linewidths. 

Kinematically split \nhhh\ profiles were detected in 
\object{Do279P12} and \object{SS3}. The velocity splitting is $\sim 1$ \kms.
In the former object, the splitting is most probably
caused by bipolar molecular outflows observed in other molecular lines. This object
has been observed in ammonia lines with the Effelsberg telescope three decades ago.
We compared these observations with the current spectra and found velocity shifts
at some positions which correspond to an acceleration of the gas flow of
$\dot{V} \la 0.03$ km~s$^{-1}$~yr$^{-1}$. 

The measured kinetic temperatures lie between 9~K and 12~K  
for the starless and isolated star-forming sources, 
except for the fast rotator \object{Do279P7} where \Tkin\ ranges between 12~K and 15~K.
An increased value of \Tkin\ in this case 
may be due to magnetic energy dissipation 
since magnetic and nonthermal energy densities 
may be nearly equal in \object{Do279P7}:
for $n = 1.1\times10^4$ \cmm, $v = 0.4$ \kms, and assuming $B \sim 20~\mu$G\footnote{In cores, 
ambipolar diffusion leads to a mass to magnetic flux ratio,
$M/\Phi = 7.6\times10^{-21}N({\rm H}_2)/B_{\rm tot}$, of 
$M/\Phi \sim 1$ (e.g., Crutcher 2012).
With the measured ammonia column density $N \sim 6\times10^{14}$ \cm\ and 
the abundance ratio [NH$_3$]/[H$_2$] $\sim 3\times10^{-7}$, one finds 
$B_{\rm tot} \sim 20~\mu$G. }
we obtain $B^2/8\pi \approx \rho v^2/2 \sim 2\times10^{-11}$ dyn~cm$^{-2}$.
However, to what extent the magnetic energy dissipation contributes to the 
gas heating requires further investigations. 

For the complex filamentary dark clouds, \Tkin\ varies
between 9~K and 18~K and strongly fluctuates from point to point. 
We found no simple relationship between gas density and kinetic temperature.
High density condensations in the filamentary dark clouds 
have a large range of temperatures, most probably
determined by radiation from embedded stars and dust. 

For the beam filling factor $\eta = 1$, we estimated the excitation temperature of an inversion
doublet, the ammonia column densities, and H$_2$ densities for those clouds where both the (1,1)
and (2,2) transitions were detected. The typical values of \Tex\ lie between 3~K and 8~K, the
total \nhhh\ column densities between $0.3\times10^{15}$ \cm\ and $3\times10^{15}$ \cm,
and H$_2$ volume densities between $0.3\times10^4$ \cmm\ and $4.2\times10^4$ \cmm.

The study of Aquila dense cores
will be continued, with the remaining $\sim 100$ targets scheduled for observations
at the Effelsberg 100-m telescope in 2013.

\begin{acknowledgements}
We thank the staff of the Effelsberg 100-m telescope for the assistance in observations and
acknowledge the help of Benjamin Winkel in preliminary data reduction. 
We also thank an anonymous referee and Malcolm Walmsley for suggestions
that led to substantial improvements of the paper. We acknowledge Dima Shalybkov
for valuable comments. SAL is grateful for the kind hospitality of 
the Max-Planck-Institut f{\"u}r Radioastronomie, Hamburger Sternwarte and Shanghai Astronomical 
Observatory where this work has been done.  SAL's work is supported by the grant DFG 
Sonderforschungsbereich SFB 676 Teilprojekt C4.
\end{acknowledgements}

\clearpage

\clearpage
\begin{table*}[t!]
\centering
\caption{Peak intensities of \nhhh(1,1) line emission toward Aquila targets 
}
\label{tbl-1}
\begin{tabular}{l c c c r@{,}l  r c r@{.}l r l}
\hline
\hline
\noalign{\smallskip}
\multicolumn{1}{c}{Source} & Peak 
& \multicolumn{2}{c}{Position}& \multicolumn{2}{c}{Offset}& 
\multicolumn{1}{c}{\Tmb$^b$} & \multicolumn{1}{c}{$V_{\scriptscriptstyle\rm LSR}$} 
& \multicolumn{2}{c}{$\Delta v^c$} & 
\multicolumn{1}{c}{Date} & \multicolumn{1}{c}{Other name} \\[-2pt]
 & \multicolumn{1}{c}{Id.$^a$} & \multicolumn{1}{c}{$\alpha_{2000}$} & \multicolumn{1}{c}{$\delta_{2000}$} 
& {$\Delta\alpha$} & {$\Delta\delta$} 
& \multicolumn{1}{c}{(K)} & \multicolumn{1}{c}{(\kms)} & \multicolumn{2}{c}{(\kms)} 
& \multicolumn{1}{c}{(d-m-y)} & \\
&  & ($^{\rm h}$: $^{\rm m}$: $^{\rm s}$) & ({\degr}: {\arcmin}: {\arcsec}) &
({\arcsec})&({\arcsec})  \\
\noalign{\smallskip}
\hline

\noalign{\smallskip}
\object{Do243~P2} & $\alpha$  & 18:04:49 & $-$04:32:38 & $40$&$0$ & 0.30(6) & 6.97(1) & 0&39(6) & 30-03-11 & 
\object{J180447.8-043208}$^1$ \\ 

\noalign{\smallskip}
\object{Do279~P6} & $\alpha$ & 18:29:07 & +00:30:51 & $0$ & $0$   
   & 2.9(6) & 7.75(1) & 0&753(7) & 26,29-03-11 & \object{Ser G3-G6}$^2$ \\[-2pt] 
& $\beta$         &             &             & $40$&$40$   & 2.1(4) & 8.04(1) & 0&63(1)& 26,29-03-11 & \\[-2pt] 
& $\gamma$        &             &             & $-160$&$-80$& 1.8(4) & 7.69(1) & 0&96(3)& 26,29-03-11 & \\[-2pt] 
& $\delta$        &             &             & $-120$&$-40$& 1.7(3) & 7.91(1) & 0&7(1)& 26,29-03-11 & \\[-2pt] 
& $\varepsilon$      &             &             & $80$&$80$& 1.3(2) & 7.81(1) & 0&4(1)& 26,29-03-11 & \\ 

\noalign{\smallskip}
\object{Do279~P12}& $\alpha$ &18:29:47 & +01:14:53 & $160$&$-40$ & 3.2(6) & 7.59(1)& 0&90(3)&  28-03-11 & 
\object{BDN~31.57+5.37}$^3$  \\[-2pt] 
& $\beta$                &             &             & $200$&$-120$& 2.8(6) & 7.09(1)& 0&84(3)& 28-03-11 & \\[-2pt]
& $\gamma$               &             &             & $40$&$40$   & 2.9(6) & 8.29(1)& 0&91(2)& 28-03-11 & \\

\noalign{\smallskip}
\object{SS3} & $\alpha$ &18:29:57 & $-$02:00:10 & $80$&$-160$ & 5(1) & 7.39(1)& 1&140(9)  &  13-01-12 & \\[-2pt]
& $\beta$        &             &             & $0$&$160$       & 3.1(6) & 7.62(1)& 0&396(4)   & 13-01-12 \\[-2pt]        
& $\gamma$       &            &             & $0$&$80$       & 3.0(6) & 7.67(1)& 0&488(6)   & 12-01-12 \\        

\noalign{\smallskip}
\object{Do279~P8} & $\alpha$ &18:30:04 & $-$02:48:14 & $0$ & $0$   & 2.5(5) & 8.52(1) & 0&26(2) & 30-03-11 & \\ 

\noalign{\smallskip}
\object{Do279~P7} & $\alpha$ &18:31:08 & $-$02:10:22 & $-80$&$40$ & 1.5(3) & 6.04(1)& 0&74(2)  &  12-01-12 & 
\object{BDN~28.65+3.66}$^3$\\

\noalign{\smallskip}
\object{Do279~P18} & $\alpha$ & 18:36:48 & $-$01:13:27 & $0$&$0$ & 0.20(4) & 8.53(1)& 1&1(1)  &  28-03-11 & 
\\[-2pt]

\noalign{\smallskip}
\object{Do279~P13} &$\alpha$ & 18:39:15 & +00:28:50 & $-40$&$0$ & 1.0(2) & 8.26(1)& 0&51(3)  &  28-03-11 & 
\\[-2pt]

\noalign{\smallskip}
\object{Do321~P1} & $\alpha$ & 18:57:10 & +00:52:58 & $-40$&$0$ & 0.15(3) & 13.6(1)& 0&30(3)  &  29-03-11 & 
\\[-2pt]
& $\beta$   &  &                                          & $0$&$40$ & 0.10(2) & 13.5(1)& 1&0(2)  &  29-03-11 & 
\\[-2pt]

\noalign{\smallskip}
\object{Do321~P2} & $\alpha$ &19:20:31 & +11:17:56 & $-80$&$120$ & 3.0(6) & 6.88(1)& 0&49(1)  &  29-03-11 & 
\object{BDN~46.22--1.34}, \\[-2pt]
& $\beta$        &             &             & $-80$&$40$ & 2.7(6)  & 7.06(1)& 0&43(1)  & 29-03-11 &  
\object{L673}, \object{P77}$^3$\\[-2pt]
& $\gamma$         &             &             & $-120$&$-40$ & 1.7(3)  & 6.93(1)& 0&38(2) &  29-03-11 & \\[-2pt]
& $\delta$         &             &             & $-80$&$240$  & 1.6(3)  & 7.12(1)& 0&47(2)  &  29-03-11 & \\[-2pt]
& $\varepsilon$       &             &             & $80$&$-40$  & 1.3(3)  & 6.66(1)& 0&73(5)  &  29-03-11 & \\

\noalign{\smallskip}
\object{Ka05} & $\alpha$ & 19:37:01 & +07:34:30 & $0$&$0$ & 2.5(5) & 8.40(1)& 0&365(8)  &  12-01-12  & 
\object{B335}$^4$, \object{CB199}$^5$\\

\noalign{\smallskip}
\object{Ka01}& $\alpha$ & 19:41:04 & +10:57:17 & $0$ & $-40$ & 2.5(5) & 9.55(1) 
& 0&311(8) & 28-03-11 & \object{B143}$^4$, \object{L694--2}$^6$ \\

\noalign{\smallskip}
\hline
\noalign{\smallskip}
\multicolumn{12}{l}{{\bf Notes.} $^a$Greek letters mark peaks of ammonia emission indicated in Figs.~\ref{Fg1}-\ref{Fg3},
\ref{Fg5}, \ref{Fg7}-\ref{Fg10}. }\\
\multicolumn{12}{l}{ $^b$The number in parentheses correspond to a $1\sigma$ statistical error at
the last digital position.  $^c$Linewidth (FWHM). }\\  
\multicolumn{12}{l}{References. (1) Di Francesco \etal\ 2008; 
(2) Cohen \& Kuhi 1979; 
(3) Dobashi \etal\ 2005; (4) Barnard 1927; } \\ 
\multicolumn{12}{l}{(5) Clemens and Barvainis 1988;  (6) Lee and Myers 1999. }
\end{tabular}
\end{table*}

\clearpage
\begin{table*}[t!]
\centering
\caption{Molecular cores toward Aquila targets
without detected \nhhh\ emission in the (1,1) and (2,2) inversion transitions 
}
\label{tbl-2}
\begin{tabular}{r l c c c r c l}
\hline
\hline
\noalign{\smallskip}
No. & Source & \multicolumn{1}{c}{rms$^a$} &
\multicolumn{2}{c}{Position}& \multicolumn{1}{c}{Date} & Mapped area & Other name \\
 &  & (K) & $\alpha_{2000}$ & $\delta_{2000}$ & \multicolumn{1}{c}{ (d-m-y) }
& $\Delta\alpha \times \Delta \delta $ \\
 &  &     & ($^{\rm h}$: $^{\rm m}$: $^{\rm s}$) & ({\degr}: {\arcmin}: {\arcsec})  & &
({\arcsec}) ({\arcsec})  \\
\noalign{\smallskip}
\hline

\noalign{\smallskip}
1 & \object{Do243 P3} & 0.24 & 18:03:07 &  $-$04:52:30 & 30-03-11 &  $80\times80$ & \\[-1pt]
2 & \object{Do243 P1} & 0.23 & 18:03:51 &  $-$04:29:33 & 29,30-03-11 &  $80\times80$ & \\[-1pt]
3 & \object{Do243 P6} & 0.22 & 18:08:43 &  $-$03:25:03 & 30-03-11 &  $80\times80$ & \\[-1pt]
4 & \object{Do243 P8} & 0.24 & 18:13:48 &  $-$03:15:20 & 30-03-11 &  $80\times80$ & \\[-1pt]
5 & \object{Do243 P7} & 0.24 & 18:16:21 &  $-$03:24:49 & 30-03-11 &  $80\times80$ & \\[-1pt]
6 & \object{Do279 P19}& 0.24 & 18:21:00 &  $-$03:01:10 & 30-03-11 &  $40\times80$ & \\[-1pt]
7 & \object{Do279 P21}& 0.24 & 18:21:45 & $-$03:07:36 & 30-03-11 &  $80\times80$ & \\[-1pt]
8 & \object{Do279 P27}& 0.22 & 18:23:43 & $-$03:14:34 & 28-03-11 &  $80\times80$ & \\[-1pt]
9 & \object{Do279 P25}& 0.26 & 18:25:34 &  $-$02:54:04 & 30-03-11 &  $80\times80$ & \\[-1pt]
10 & \object{Do279 P2} & 0.19 & 18:26:59 & $-$03:43:44 & 28-03-11 &  $80\times80$ & \\[-1pt]
11 & \object{Do279 P3} & 0.23 & 18:28:21 &  $-$03:29:20 & 30-03-11 &  $80\times80$ & \\[-1pt]
12 & \object{Do279 P5}& 0.24 &18:28:42 &  $-$03:44:58 & 30-03-11 &  $80\times80$ & \\[-1pt]
13 &\object{Do279 P14}& 0.26 & 18:29:15 &  $-$03:19:38 & 30-03-11 &  $80\times80$ & \\[-1pt]
14 &\object{Do279 P22} & 0.24 & 18:29:18 & $-$01:25:33 & 30-03-11 &  $80\times80$ & \\[-1pt]
15 &\object{Do279 P4} & 0.24 & 18:30:09 &  $-$03:42:28 & 30-03-11 &  $80\times80$ & \\[-1pt]
16 &\object{Do279 P1} & 0.24 & 18:31:31 &  $-$02:30:29 & 30-03-11 &  $80\times80$ & \\[-1pt]
17 &\object{Do279 P24}& 0.22 & 18:33:20 &  +00:42:42 & 30-03-11 &  $80\times80$ & 
\object{LND583}$^1$\\[-1pt]
18 &\object{Do279 P17} & 0.24 & 18:37:41 & +00:13:06 & 28-03-11 &  $80\times80$ & \\[-1pt]
19 &\object{Do279 P20} & 0.24 & 18:38:29 & $-$00:02:30 & 30-03-11 &  $80\times80$ & \\[-1pt]
20 &\object{Do296 P3}  & 0.18 & 18:51:21 & $-$04:16:45 & 28-03-11 &  $80\times80$ & \\[-1pt]
21 & \object{Do321 P8} & 0.22 & 18:55:36 & +02:12:00 & 28-03-11 &  $80\times80$ & \\[-1pt]
22 & \object{Do321 P5}& 0.22 & 19:01:38 &  +04:05:11 & 30-03-11 &  $80\times80$ & \\[-1pt]
23 & \object{Do299 P1} & 0.19 & 19:07:31 & $-$03:55:48 & 29-03-11 &  $80\times80$ & 
\object{B135}$^2$\\[-1pt]
24 & \object{Do321 P7}& 0.23 & 19:10:48 &  +07:26:14 & 30-03-11 &  $80\times80$ & \\[-1pt]
25 & \object{Ka09}    & 0.20 & 19:35:11 &  +12:25:40 & 30-03-11 &  $80\times80$ & \\[-1pt]
26 & \object{Ka03}    & 0.21 & 19:40:36 &  +10:20:14 & 30-03-11 &  $80\times80$ & 
\object{LDN690}$^1$\\[-1pt]
27 & \object{Ka16}     & 0.15 & 19:43:08 & $-$05:42:04 & 28,30-03-11 &  $320\times320$ & \\[-1pt]
28 & \object{Ka19}    & 0.26 & 19:43:15 &  $-$07:13:44 & 30-03-11 &  $80\times80$ & \\[-1pt]
29 & \object{Ka02}    & 0.23 & 19:44:10 &  +10:28:28 & 30-03-11 &  $80\times80$ & \\[-1pt]
30 & \object{Ka21}     & 0.19 & 19:45:11 &  $-$07:07:41 & 28-03-11 &  $80\times80$ & \\[-1pt]
31 & \object{Ka06}     & 0.17 & 19:47:50 &  +10:50:16 & 28-03-11 &  $80\times80$ & \\[-1pt]
32 & \object{Ka07}     & 0.15 & 19:48:48 &  +11:23:07 & 28-03-11 &  $80\times80$ & 
\object{B340}$^2$\\[-1pt]
33 & \object{Ka15}    & 0.27 & 19:48:48 &  $-$10:30:46 & 30-03-11 &  $80\times80$ & \\[-1pt]
34 & \object{Ka17}    & 0.23 & 19:49:44 &  $-$06:21:44 & 30-03-11 &  $80\times80$ & \\[-1pt]
35 & \object{Ka14-15-A}& 0.30 & 19:51:49 & $-$10:34:32 & 30-03-11 &  $120\times80$ & \\[-1pt]
36 & \object{Ka14}     & 0.32 & 19:51:58 & $-$10:32:12 & 29,30-03-11 &  $80\times80$ & \\[-1pt]
37 & \object{Ka23}    & 0.27 & 19:58:36 &  $-$14:06:00 & 30-03-11 &  $80\times80$ & 
\object{MBM159}$^3$\\[-1pt]

\noalign{\smallskip}
\hline
\noalign{\smallskip}
\multicolumn{8}{l}{{\bf Notes.}\ $^a$Mean noise level per channel width of 0.077 \kms\ on a \Tmb\ scale.}\\
\multicolumn{8}{l}{References. (1) Lynds 1962; (2) Barnard 1927; (3) Magnani \etal\ 1985. }
\end{tabular}
\end{table*}

\clearpage
\begin{appendix} 
\section{The analysis of \nhhh(1,1) and (2,2) spectra}

In this Appendix we discuss the methods used for calculating physical parameters from
observations of the ammonia \nhhh\ $(J,K) = (1,1)$ and (2,2) transitions. 
We give a consistent summary of the basic formulae relevant to the 
study of cold molecular cores. 

The recorded \nhhh\ spectra were first converted into the \Tmb\ scale
and a baseline was removed from each of them. 
The baseline was typically linear in the range covered by the ammonia 
hyperfine (hf) structure lines
but occasionally quadratic.
After that the spectra of the same offset were co-added with weights
inversely proportional to the mean variance of the noise per channel, rms$^{-2}$,
estimated from the channels with no emission.

The \nhhh\ spectra were fitted to determine the
total optical depth $\tau_{\rm tot}$ in the respective inversion transition,
the LSR velocity of the line \VLSR, 
the intrinsic full width to half power linewidth \Dv\ for an individual hf 
component,  and the amplitude, $\cal{A}$ (e.g., Ungerechts \etal\ 1980):
\begin{equation}
T(v) = {\cal A}\left[1 - {\rm e}^{-\tau(v)}\right]\, .
\label{Eq1}
\end{equation}
The optical depth $\tau(v)$ at a given radial velocity $v$ is defined by
\begin{equation}
\tau(v) = \tau_{\rm tot} \sum^n_{i=1} r_i \exp \{ 
-2.77 [(v - V_{\scriptscriptstyle\rm LSR}) + v_i]^2 / (\Delta v)^2 \}.
\label{Eq2}
\end{equation}
In this equation, the sum runs over the $n$ hf components of the inversion transition
($n = 18$ and 21 for the \nhhh(1,1) and (2,2) inversion transitions, respectively), 
$r_i$ is the theoretical relative intensity of the $i$th hf line and $v_i$ is its
velocity separation from the fiducial frequency. The values of these parameters 
are given in, e.g., Kukolich (1967) and Rydbeck \etal\ (1977).

In Eq.(\ref{Eq2}), the total optical depth, $\tau_{\rm tot}$, is the maximum optical
depth which an unsplit (1,1) or (2,2) line would have 
at the central frequency if the hf levels 
were populated at the same excitation temperature for both lines (1,1) and (2,2). 
Assuming that the line profile function has a Gaussian shape with the
width \Dv\ (FWHM) and taking into account that the
statistical weights of the upper and lower levels of an inversion transitions are
equal, one obtains
\begin{eqnarray}
\tau_{\rm tot}(J,K) & = & \frac{16\pi^3\mu^2}{3 h \Delta v} \left(
\frac{\ln 2}{\pi}\right)^{1/2} \frac{K^2}{J(J+1)} N(J,K) \cdot \nonumber\\ 
 &  & \cdot 
\frac{1-\exp(-h\nu/k_{\scriptscriptstyle\rm B}T_{\rm ex})}{1+\exp(-h\nu/k_{\scriptscriptstyle\rm B}T_{\rm ex})}\, ,
\label{Eq3}
\end{eqnarray}
where $h$ and $k_{\scriptscriptstyle\rm B}$ are Planck's and Boltzmann's constants, respectively,
$\mu$ is the dipole moment, $|\mu| = 1.4719$ Debye (JPL catalog\footnote{http://spec.jpl.nasa.gov/}),  
$N(J,K)$ is the inversion state column density~--- the sum of the column densities 
of the upper and lower levels of an inversion doublet,
$\nu$ is the rest frequency of the inversion transition,
and \Tex\ is the excitation temperature
which characterizes the \nhhh\ population across the $(J,K)$ inversion doublet.

In Eq.(\ref{Eq1}), the amplitude ${\cal A}$ can be expressed in terms of the beam filling factor,
$\eta = \Omega_{\rm cloud}/\Omega_{\rm beam}$ (solid angle $\Omega = \pi \theta^2 /(4\ln 2)$, 
where $\theta$ is the angular diameter at FWHP), 
and the excitation temperature, $T_{\rm ex}$, as
\begin{equation}
{\cal A} = \eta \left[ J_\nu(T_{\rm ex}) - J_\nu(T_{\rm bg}) \right]\, ,
\label{Eq4}
\end{equation}
where $T_{\rm bg} = 2.7$~K is the black body background
radiation temperature, and $J_\nu(T)$ is defined by
\begin{equation}
J_\nu(T) = \frac{h\nu}{k_{\scriptscriptstyle\rm B}}\left( {\rm e}^{h\nu/k_{\scriptscriptstyle\rm B}T} -1 \right)^{-1}.
\label{Eq5}
\end{equation}

For optically thin lines (e.g., $\tau_{11} \ll 1$), Eq.(\ref{Eq1}) degenerates into
\begin{eqnarray}
T(v) & =& {\cal A}\cdot\tau(v) \nonumber\\
     & =& ({\cal A}\tau_{11}) \cdot \sum^n_{i=1} r_i \exp [ 
-2.77 (\Delta V_i / \Delta v)^2 ]\, ,
\label{Eq7}
\end{eqnarray}
and it is not possible to determine ${\cal A}$ and $\tau_{11}$ separately.
Here $\Delta V_i = v - V_{\scriptscriptstyle\rm LSR} + v_i$. 

For some molecular cores we observe two-component \nhhh\ profiles. The line parameters 
$\{ {\cal A}_j, \tau^{(j)}_{\rm 11}, \tau^{(j)}_{\rm 22}, V^{(j)}_{\scriptscriptstyle\rm LSR}, \Delta v_j \}$ 
for both $j=1$ and $j=2$ components 
were determined in this case by fitting the function
\begin{equation}
T(v) = T_1(v) + T_2(v)\, .
\label{Eq8}
\end{equation}
This linear form assumes 
that two clouds in the beam have the same filling factor $\eta$
and the same \Tex\ for both lines (1,1) and (2,2). 

At low spectral resolutions, when the channel spacing, $\Delta_{\rm ch}$, 
is comparable to the linewidth \Dv, 
the derived line parameters are affected by the channel profile
$\phi(v)$. In this case, the recorded spectrum, $T'(v)$, was considered as 
a convolution of the true spectrum $T(v)$ and $\phi(v)$ 
\begin{equation}
T'(v) = \int T(v)\cdot\phi(v - v')\, dv'\, .
\label{Eq9}
\end{equation}
We approximate $\phi(v)$ by a Gaussian function
with the width FWHM defined by a particular backend setting.

The calculated synthetic spectrum which is defined as
$T_{\rm syn}(v) \equiv T(v)$ if $\Delta_{\rm ch} \ll \Delta v$, 
or $T_{\rm syn}(v) \equiv  T'(v)$ if $\Delta_{\rm ch} \sim \Delta v$,
was fitted to 
the observed spectrum $T_{\rm obs}(v)$ through minimization of
the $\chi^2$ function
\begin{equation}
\chi^2 = \sum\, \left[ T_{\rm syn}(v) - T_{\rm obs}(v) \right]^2/{\rm rms}^2
\label{Eq10}
\end{equation}
in the 5-dimensional space of the following model parameters:
${\cal A}, \tau_{\rm 11}, \tau_{\rm 22},
V_{\scriptscriptstyle\rm LSR},$ and \Dv. 
Attempting to exclude at the $3\sigma$ level 
all noise-only channels 
in between the hf lines, the sum runs
over the channels with observed \nhhh\ emission from all hf components.
The minimum of $\chi^2$ was computed using the simplex method (e.g., Press  \etal\ 1992). 

Equation~(\ref{Eq10}) was also used to estimate the formal errors of the model
parameters through calculation of the covariance matrix at the minimum of $\chi^2$.
Since the uncertainty in the amplitude scale calibration was $\sim $15-20\% (Sect.~\ref{sect-2}), 
and there were no noticeable correlations between the sequential channels in our datasets,
we did not correct the calculated errors by an additional factor as described, e.g., in 
Rosolowsky \etal\ (2008). 

Given the estimate of the amplitude ${\cal A}$, the excitation temperature $T_{\rm ex}$
can be calculated from Eq.(\ref{Eq4}):
\begin{equation}
T_{\rm ex} = \frac{ T_0 }  
{ \ln \left[ 1 +  T_0 \eta / ( {\cal A} + \eta J_{\rm bg} ) \right] }   \, ,
\label{Eq11}
\end{equation}
where $T_0 = h\nu/k{\scriptscriptstyle\rm B}$,  and $J_{\rm bg} = J_\nu(T_{\rm bg})$. 
At $T_{\rm bg} = 2.725$~K and \nhhh-inversion frequencies, we have
$J_{\rm bg} = 2.20$~K and  $T_0 = 1.14$~K. 

In this equation the only unknown parameter is the filling factor, $\eta$. 
If the source is completely resolved, then $\eta = 1$. 
Otherwise, Eq.~(\ref{Eq11}) gives a lower bound on \Tex\ corresponding to $\eta = 1$,
whereas decreasing $\eta$ drives \Tex\ towards its upper bound at \Tex\ = \Tkin,
which holds for LTE. 
In case of emission, the radiation temperature within the line is larger than the 2.7~K
background and the excitation may be either radiative or collisional. 
At gas densities larger than the critical density\footnote{For \nhhh(1,1), 
$n_{\rm cr} = 3.90\times10^3$ \cmm\ at \Tkin\ = 10~K (Maret \etal\ 2009).}
and for optically thin lines the local density
of line photons is negligible as compared to the background radiation field,
and the lower metastable states in \nhhh\ are mainly populated via collisions with
molecular hydrogen\footnote{The abundance ratio [H$_2$]/[He] $\sim 5$ and
the \nhhh-H$_2$ collision rates are a factor of 3 larger 
than the \nhhh-He rates (Machin \& Roueff 2005).}, 
i.e., the stimulated emission can be neglected.
For low collisional excitation, $n_{\rm gas} < n_{\rm cr}$, the line radiation 
controls \Tex\ for both optically thin and thick regimes (Kegel 1976). 
\nhhh(1,1) emission usually arises from regions with $n_{\rm gas} \ga 10^4$ \cmm\ 
(e.g., Ho \& Townes 1983).
It means that the regime of low collisional excitation does not dominate. 
We also do not consider anomalous excitation (maser emission), leading to \Tex\ $\gg T_{\rm bg}$, 
since masing is not observed in our data sample. 
Thus, the use of \Tex\ = \Tkin\ as a formal upper limit for
the excitation temperature is legitimate. 
By this we can restrict the unknown filling factor for an unresolved source in the interval
\begin{equation}
\eta_{\rm min} \leq \eta \leq \eta_{\rm max}\ , 
\label{Eq11a}
\end{equation}
and \Tex\ within the boundaries
\begin{equation}
\left. T^{\rm min}_{\rm ex} \right|_{\eta = 1} \leq T_{\rm ex} \leq 
\left. T^{\rm max}_{\rm ex} \right|_{\eta=\eta_{\rm min}} \ .
\label{Eq11b}
\end{equation}

From a separate analysis of the (1,1) and (2,2) lines with low noise level, we found
that both amplitudes ${\cal A}_{11}$ and ${\cal A}_{22}$ 
are equal within the observational errors, 
which implies that $T_{\rm ex}(1,1) \approx T_{\rm ex}(2,2)$. 
Albeit radiative transfer
calculations show that the excitation temperatures $T_{\rm ex}(1,1)$ and  $T_{\rm ex}(2,2)$
may differ by about 20\% 
(Stutzki \& Winnewisser 1985), the assumption of their equivalence seems
to be sufficiently good in our case.

The second temperature which describes the \nhhh\ population is the rotational temperature,
\Trot, characterizing the population of energy levels with different $(J,K)$.
As mentioned above, the population 
of the lower metastable inversion doublets is determined by collisions with H$_2$ and,
thus, is regulated by the kinetic temperature, \Tkin.
The population ratio between the (1,1) and (2,2) states is defined as
\begin{equation}
\frac{N_{22}}{N_{11}} = \frac{g_{22}}{g_{11}}\cdot\exp\left( - \frac{\Delta E_{12}}{
k_{\scriptscriptstyle \rm B} T_{\rm rot} } \right) =
\frac{5}{3}\cdot\exp\left( - \frac{41.5 {\rm K}} {T_{\rm rot} } \right).
\label{Eq13}
\end{equation}
Assuming that both transitions trace the same volume of gas and their linewidths \Dv\ 
are equal, one finds from Eqs.(\ref{Eq3}) and (\ref{Eq13}) that
\begin{equation}
T_{\rm rot}  = - 41.5 / \ln \left( \frac{9}{20} \cdot 
\frac{\tau_{22}}{\tau_{11}} \right) \, . 
\label{Eq14}
\end{equation}
We note that since radiative transitions
between the different $K$-ladders of the $(J,K)$ levels are forbidden
and that the unknown filling factor $\eta$ is canceled in the line
intensity ratio of the \nhhh\ (1,1) and (2,2) transitions,
the estimate of the rotational temperature is 
more reliable than the excitation temperature. 

For a two-level system  
with the kinetic temperature less than the energy gap between the (1,1) and (2,2) states
($\Delta E_{12} = 41.5$ K), the rotational temperature can be related to the kinetic
temperature through detailed balance arguments (e.g., Ragan \etal\ 2011): 
\begin{equation}
T_{\rm kin}  = T_{\rm rot} / [ 1 - \frac{ T_{\rm rot} }{ \Delta E_{12} }\cdot
\ln (1 + 1.1\ {\rm e}^{-15.7/T_{\rm rot}})]\ .
\label{Eq18}
\end{equation}

The beam-averaged column density $N_{\scriptscriptstyle JK}$ 
(in \cm) can be calculated from Eqs.(\ref{Eq3}) and (\ref{Eq4}) 
using the estimated values of the amplitude ${\cal A}$, the total optical 
depth $\tau_{\scriptscriptstyle JK}$, and  the linewidth \Dv :
\begin{equation}
N_{\scriptscriptstyle JK} = 
\zeta_{\scriptscriptstyle JK}\cdot \Delta v \cdot \tau_{\scriptscriptstyle JK} \cdot 
\frac{{\cal A} + \eta (J_{\rm bg} + 0.5T_0)}{\eta}\ ,
\label{Eq15}
\end{equation}
where $\zeta_{\scriptscriptstyle 11} = 1.3850\times10^{13}$,
$\zeta_{\scriptscriptstyle 22} = 1.0375\times10^{13}$,
and \Dv\ is in \kms\ (for detail, see  Ungerechts \etal\ 1980).
As mentioned above, ${\cal A}_{11} \approx {\cal A}_{22}$, and
for both transitions $T_0 = 1.14$~K, and $J_{\rm bg} = 2.20$~K. 
Substituting these numerical values in Eq.(\ref{Eq15}), one obtains 
\begin{equation}
N_{11} = 3.84\times10^{13} \cdot \Delta v \cdot \tau_{\scriptscriptstyle 11} \cdot 
(1 + {\cal A}/2.77\eta)\ .
\label{Eq15a}
\end{equation}
This gives us lower and upper boundaries of $N_{11}$ 
if $\eta \in [\eta_{\rm min}, 1]$\ : 
\begin{equation}
\left. N^{\rm min}_{11}\right|_{\eta=1} \leq N_{11} \leq  
\left. N^{\rm max}_{11}\right|_{\eta=\eta_{\rm min}} .
\label{Eq15b}
\end{equation}

The uncertainty interval $\Delta N_{11} = N^{\rm max}_{11} - N^{\rm min}_{11}$ transforms
directly into the estimate of the total \nhhh\ column density:
\begin{eqnarray}
N({\rm NH}_3) & = & N_{11}\cdot [(1/3)\exp(23.2/T_{\rm rot}) + 1 + \nonumber\\ 
       &  &  + (5/3)\exp(-41.5/T_{\rm rot}) + \nonumber\\
       &  &  + (14/3)\exp(-105.2/T_{\rm rot}) + \ldots ]\ ,
\label{Eq16}
\end{eqnarray}
which assumes that the relative population of all metastable levels of both ortho-\nhhh\
($K = 3$), which is not observable,
and para-\nhhh\   ($K = 1, 2$)
is governed by the rotational temperature of the system at thermal equilibrium
(Winnewisser \etal\ 1979). 
Substituting $N^{\rm min}_{11}$ and $N^{\rm max}_{11}$ from (\ref{Eq15b}) into (\ref{Eq16}), one finds: 
\begin{equation}
\left. N({\rm NH}_3)_{\rm min}\right|_{\eta=1} \leq N({\rm NH}_3) \leq  
\left. N({\rm NH}_3)_{\rm max}\right|_{\eta=\eta_{\rm min}} .
\label{Eq17}
\end{equation}

The detailed balance calculations also provide a relation between the gas density,
gas kinetic temperature, and the excitation temperature (e.g., Ho \& Townes 1983):
\begin{equation}
n({\rm H}_2) = \frac{A}{C}  \left[ \frac{ J_\nu(T_{\rm ex}) - J_\nu(T_{\rm bg}) }
{ J_\nu(T_{\rm kin}) - J_\nu(T_{\rm ex}) } \right]  \left[ 1 + 
\frac{J_\nu(T_{\rm kin})}{h\nu/k_{\scriptscriptstyle\rm B}} \right],
\label{Eq19}
\end{equation}
where $A$ is the Einstein A-coefficient, and $C$ is the rate coefficient for collisional de-excitation.
For a typical kinetic temperature in the dense molecular cores of $\sim 10-20$~K, 
the collision coefficient is 
$\sim 4\times10^{-7} {\rm s}^{-1}\ [n({\rm H}_2)/10^4 {\rm cm}^{-3}]$.
The value of $A$  for the inversion transition (1,1) is $1.67\times10^{-7}$ s$^{-1}$. 
However, the gas density calculated using Eq.(\ref{Eq19}) may be significantly underestimated
if the beam is not filled uniformly. Moreover, in case of \Tex\ = \Tkin, 
Eq.(\ref{Eq19}) is invalid and, hence, $n({\rm H}_2)$
is to be calculated by different methods
(see, e.g., Hildebrand 1983; Pandian \etal\ 2012). 
We use (\ref{Eq19}) just to set a lower bound on the gas density, $n({\rm H}_2)_{\rm min}$,
at $\eta = 1$. 

Given a fractional \nhhh\ abundance, an upper bound on the gas density may be estimated from 
the deduced $N$(NH$_3)_{\rm max}$ assuming that the ammonia emission traces the real 
distribution of the gas density\footnote{Ammonia and
submillimeter maps of dense cores in Ophiuchus show 
a close correlation between the large-scale distributions of the \nhhh(1,1) 
integrated intensity and the 850 $\mu$m continuum emission  (Friesen \etal\ 2009).}.
If the source is unresolved, its diameter, $d$, and the beam filling factor are related as
\begin{equation}
d = \theta_{\rm m} D \sqrt{\eta}\ ,
\label{Eq20}
\end{equation}
where $\theta_{\rm m}$ is the beam angular diameters (FWHP),
and $D$ the distance of the source.
The highest gas density is obtained at the smallest diameter,
$d_{\rm min} = \theta_{\rm m} D \sqrt{\eta_{\rm min}}$\ :
\begin{equation}
n({\rm H}_2)_{\rm max} = \left. \frac{N({\rm NH}_3)_{\rm max}}{X \cdot d_{\rm min}} \right|_{\eta=\eta_{\rm min}} , 
\label{Eq21}
\end{equation}
where $X = [{\rm NH}_3]/[{\rm H}_2]$ is a given abundance ratio.
In this equation the unknown distance $D$ can be found from the requirement that
both values of the lowest gas densities calculated at $\eta = 1$ from Eq.(\ref{Eq19})
and from $N({\rm NH}_3)_{\rm min}$ at $d_{\rm max} = \theta_{\rm m} D$ be equal.
This gives
\begin{equation}
n({\rm H}_2)_{\rm max} = \frac{n({\rm H}_2)}{\sqrt{\eta_{\rm min}}}
\frac{N({\rm NH}_3)_{\rm max}}{N({\rm NH}_3)_{\rm min}} \ .
\label{Eq21a}
\end{equation}

For the unresolved source its axis~--- major, $\theta_{\rm a}$, or minor, $\theta_{\rm b}$~---  
is less than $\theta_{\rm m}$, $\eta < 1$, and only limiting values can be obtained for
\Tex, $N_{JK}$, $N({\rm NH}_3)$, and $n({\rm H}_2)$.
If the source is resolved, i.e.,
$\theta_{\rm b} > \theta_{\rm m}$, then $\eta = 1$ and these physical parameters are directly defined.
In the latter case we calculate the deconvolved values of $\theta_{\rm a}$ and $\theta_{\rm b}$
and their geometrical mean 
\begin{equation}
\theta = \sqrt{\theta_{\rm a}\cdot \theta_{\rm b}}\ ,
\label{Eq22}
\end{equation}
which is used as a formal estimate of the source angular diameter.

A similar procedure is applied to calculate the virial mass of the ammonia core
through the linewidth, \Dv. If the observed value \Dv\ is of order of the
spectral resolution $\Delta_{\rm sp}$ (FWHM), then only an upper limit on 
$M_{\rm vir}$ is defined. Otherwise, if \Dv\ $> \Delta_{\rm sp}$, then the deconvolved
value of the linewidth, $\Delta \tilde{v}$, 
and the core radius $r$ give the virial mass (e.g., Lemme \etal\ 1996) 
\begin{equation}
M_{\rm vir} = 250\cdot \Delta \tilde{v}^2\cdot r\ ,
\label{Eq23}
\end{equation}
where $\Delta \tilde{v}$ is in \kms, $r$ in pc,
and $M_{\rm vir}$ in solar masses $M_\odot$. 

To conclude, we note that 
in some cases, when \Tkin\ varies slowly within an ammonia clump, 
the \nhhh\ column density can be deduced from the (1,1) line alone assuming the
same average \Tkin\ in the core as in the outskirts of a cloud. This allows us to extend
the radial gas distribution to zones with unobservable (2,2) emission (Morgan \etal\ 2013). 

\end{appendix}

\clearpage
\Online
\begin{appendix} 
\section{Ammonia spectra toward the Aquila rift cloud complex and derived physical parameters}

\begin{figure}[b]
\vspace{-15.0cm}
\hspace{0.0cm}\psfig{figure=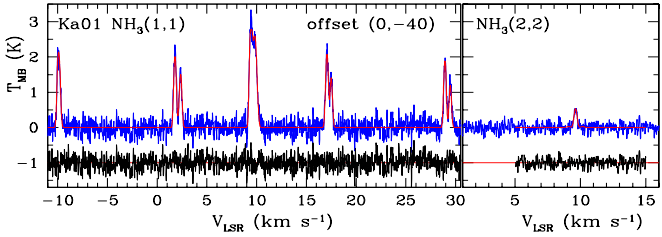,height=8.0cm,width=9.0cm}
\vspace{-5.0cm}
\caption[]{
Ammonia (1,1) and (2,2) spectra (blue)
toward the source \object{Ka01}. 
The channel spacing is 0.015 \kms, the spectral resolution FWHM = 0.024 \kms. The red curves show
the fit of a single-component Gaussian model
to the original data. 
The residuals between the observed and model spectra are shown by black color.
}
\label{afg1}
\end{figure}

\begin{figure}
\vspace{0.0cm}
\hspace{0.0cm}\psfig{figure=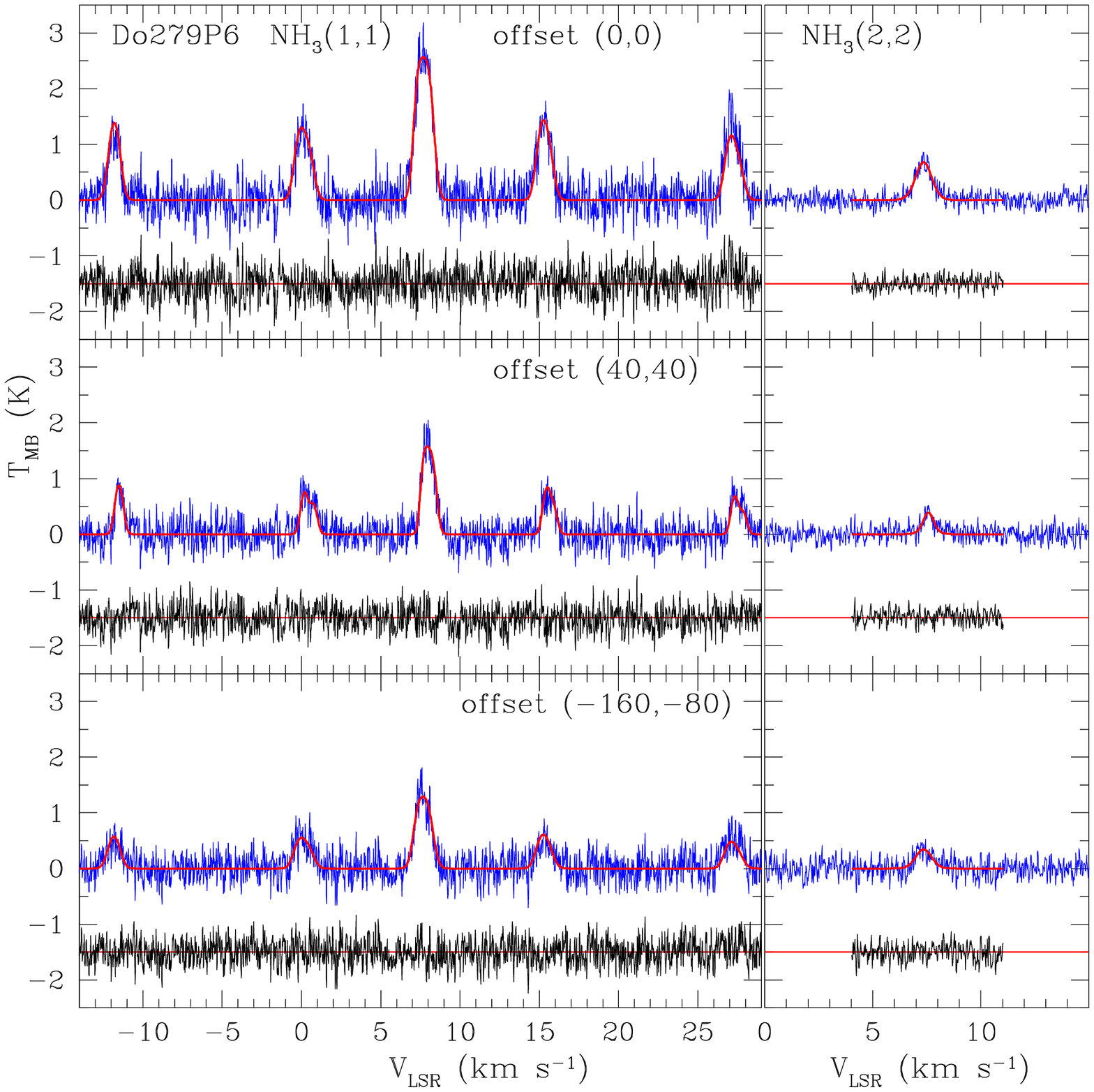,height=8.0cm,width=9.0cm}
\vspace{0.0cm}
\caption[]{
Same as Fig.~\ref{afg1} but for the source \object{Do279P6}. 
}
\label{afg2}
\end{figure}

\begin{figure}
\vspace{0.0cm}
\hspace{0.0cm}\psfig{figure=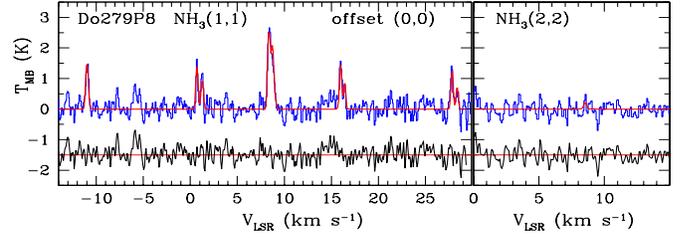,height=8.0cm,width=9.0cm}
\vspace{-5.0cm}
\caption[]{
Same as Fig.~\ref{afg1} but for the source \object{Do279P8} and
the channel spacing 0.077 \kms\ (FWHM = 0.123 \kms). 
The red curves show the fit of a single-component Gaussian model
to the \nhhh(1,1) original data and the upper limit on \nhhh(2,2).
}
\label{afg3}
\end{figure}

\begin{figure}
\vspace{0.0cm}
\hspace{0.0cm}\psfig{figure=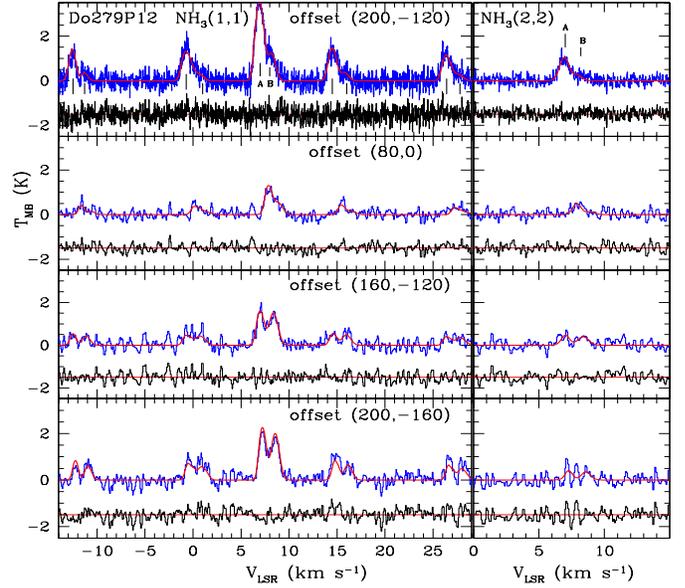,height=8.0cm,width=9.0cm}
\vspace{0.0cm}
\caption[]{
Same as Fig.~\ref{afg1} but for the source \object{Do279P12}.
The ammonia spectra have double structure. 
The upper panels show the high-resolution spectra 
(channel spacing 0.015 \kms), and the other panels~--- low-resolution spectra
(channel spacing  0.077 \kms).
Two \nhhh\ components  
are marked by ticks and labeled by letters A and B in the upper panels.
}
\label{afg4}
\end{figure}

\begin{figure}
\vspace{0.0cm}
\hspace{0.0cm}\psfig{figure=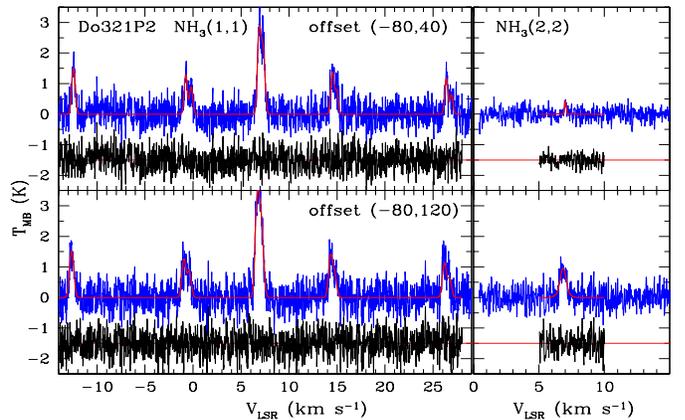,height=8.0cm,width=9.0cm}
\vspace{-2.5cm}
\caption[]{Same as Fig.~\ref{afg1} but for the source \object{Do321P2}. }
\label{afg5}
\end{figure}

\begin{figure}
\vspace{0.0cm}
\hspace{0.0cm}\psfig{figure=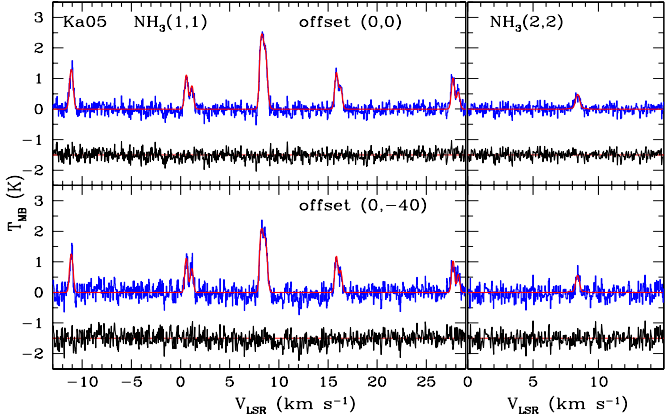,height=8.0cm,width=9.0cm}
\vspace{-2.5cm}
\caption[]{
Same as Fig.~\ref{afg1} but for the source \object{Ka05} and the
channel spacing 0.038 \kms\ (FWHM = 0.044 \kms). 
}
\label{afg6}
\end{figure}

\begin{figure}
\vspace{0.0cm}
\hspace{0.0cm}\psfig{figure=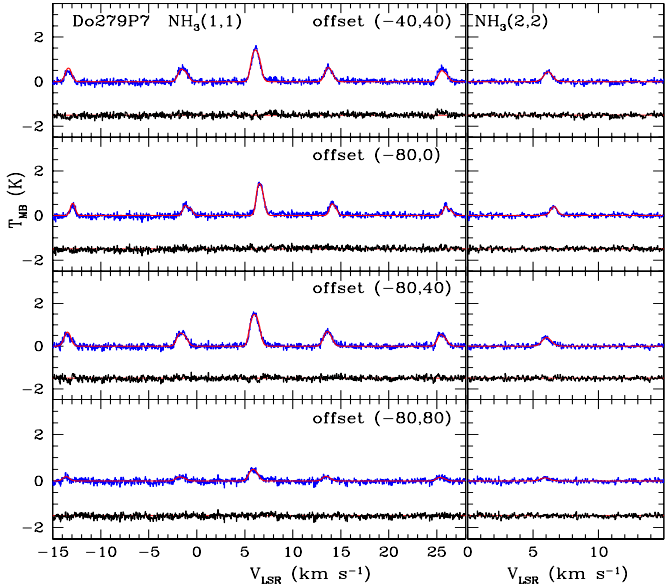,height=8.0cm,width=9.0cm}
\vspace{0.0cm}
\caption[]{
Same as Fig.~\ref{afg6} but for the source \object{Do279P7}.
}
\label{afg7}
\end{figure}

\begin{figure}
\vspace{0.0cm}
\hspace{0.0cm}\psfig{figure=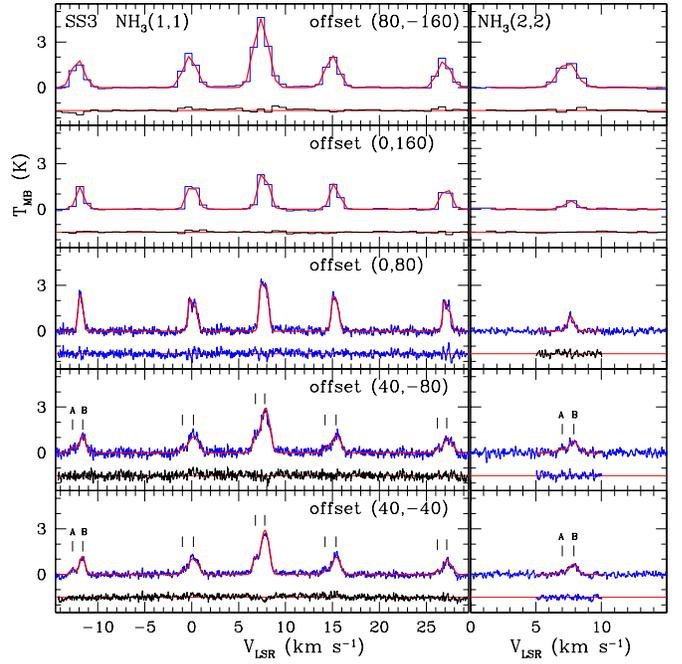,height=9.0cm,width=9.0cm}
\vspace{0.0cm}
\caption[]{
Same as Fig.~\ref{afg1} but for the source \object{SS3}. 
The channel spacing is 0.77 \kms\ and
0.038 \kms\ at two upper and three lower panels, respectively.
The corresponding spectral resolutions are 0.895 \kms\ and 0.044 \kms\ (FWHM). 
The red curves show the fit of a single-component (two upper panels) and a double-component 
(three lower panels) Gaussian model 
to the original \nhhh\ data.
Two components of the \nhhh\ emission 
are marked by ticks and labeled by letters A and B.
}
\label{afg8}
\end{figure}

\begin{figure}
\vspace{0.0cm}
\hspace{0.5cm}\psfig{figure=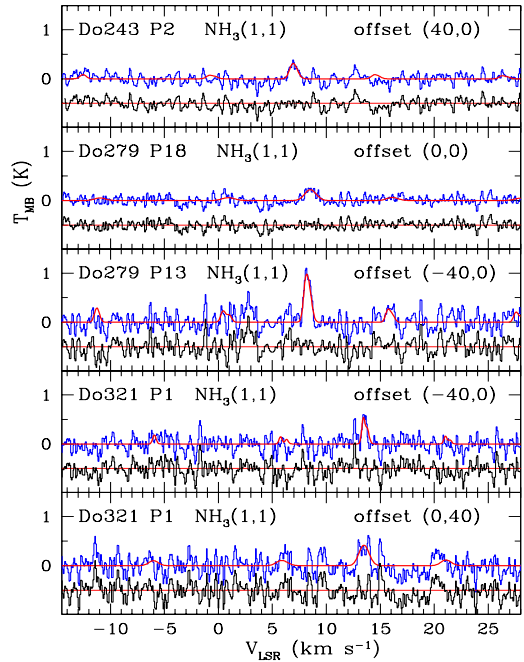,height=9.0cm,width=10.0cm}
\vspace{-0.2cm}
\caption[]{
Same as Fig.~\ref{afg3} but for the sources 
\object{Do243P2}, \object{Do279P18}, \object{Do279P13}, and \object{Do321P1}. 
}
\label{afg9}
\end{figure}

\begin{figure*}
\vspace{0.0cm}
\hspace{0.5cm}\psfig{figure=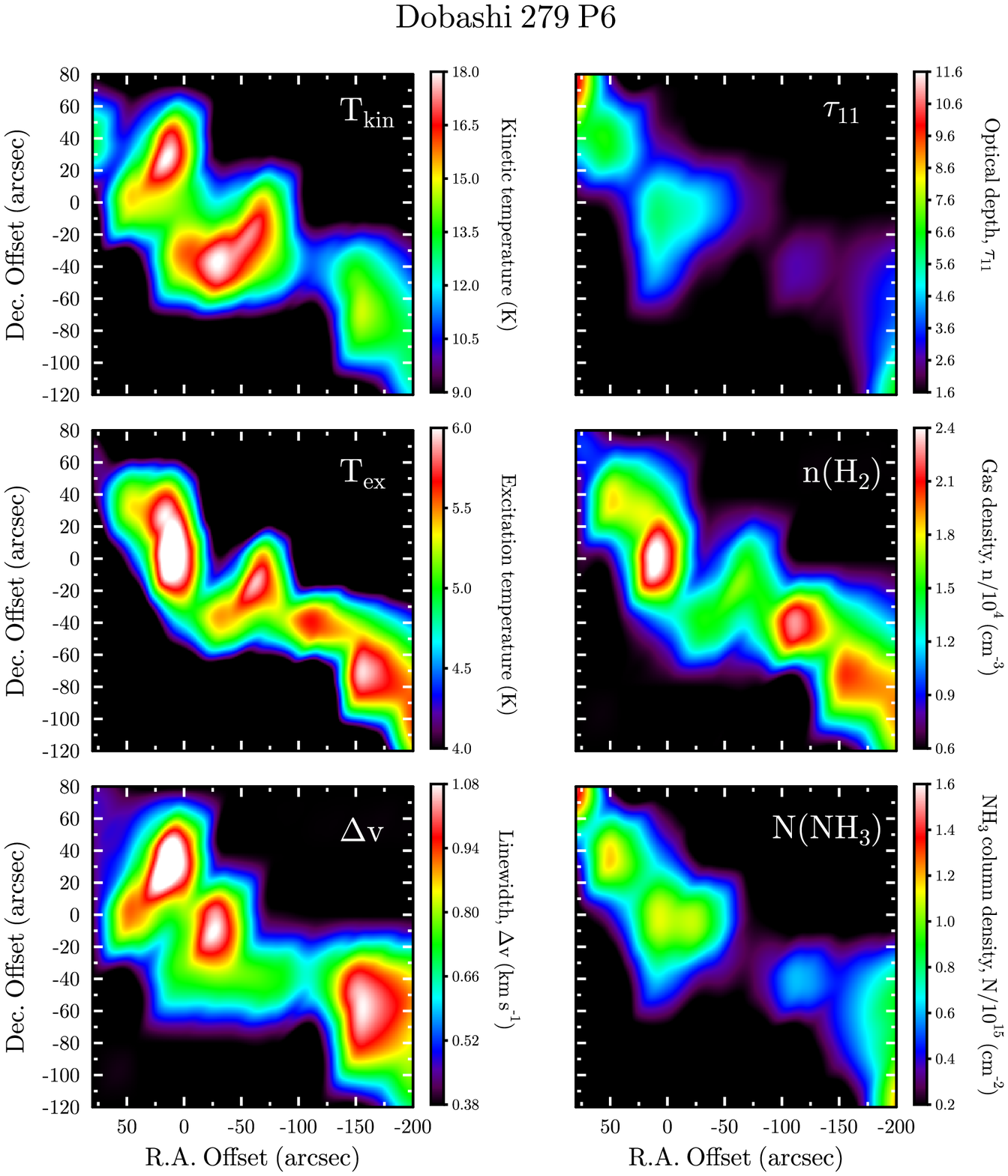,height=18.0cm,width=16.0cm}
\vspace{-2.0cm}
\caption[]{
The spatial distributions of the physical parameters 
measured in \object{Do279P6} from the ammonia inversion lines \nhhh(1,1) and (2,2).
The corresponding numerical values are listed in Table~\ref{tbl-4}.
}
\label{afg10}
\end{figure*}

\begin{figure*}
\vspace{0.0cm}
\hspace{0.5cm}\psfig{figure=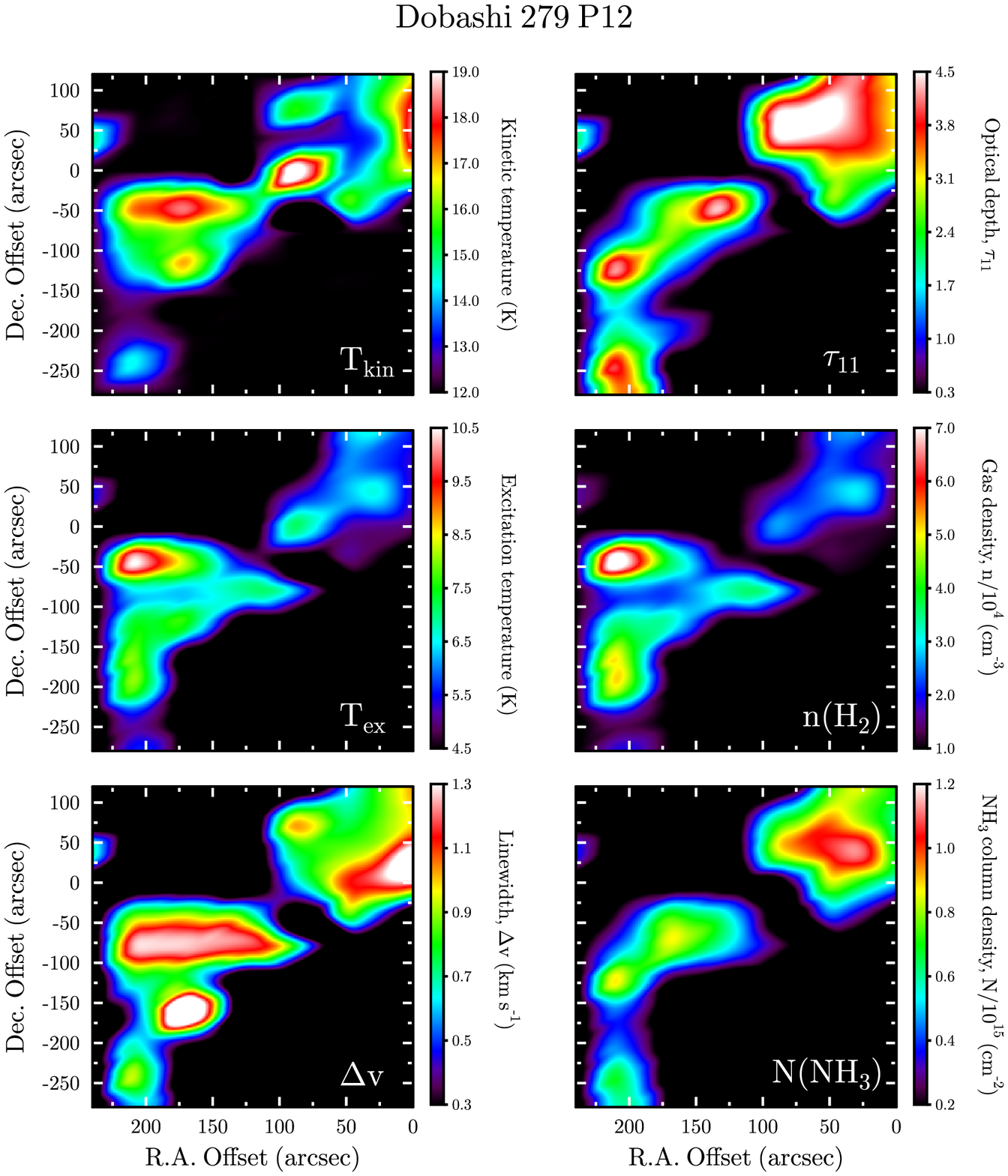,height=18.0cm,width=16.0cm}
\vspace{-2.0cm}
\caption[]{
The spatial distributions of the physical parameters 
measured in \object{Do279P12} from the ammonia inversion lines \nhhh(1,1) and (2,2).
The corresponding numerical values are listed in Table~\ref{tbl-5}.
}
\label{afg11}
\end{figure*}

\begin{figure*}
\vspace{0.0cm}
\hspace{0.5cm}\psfig{figure=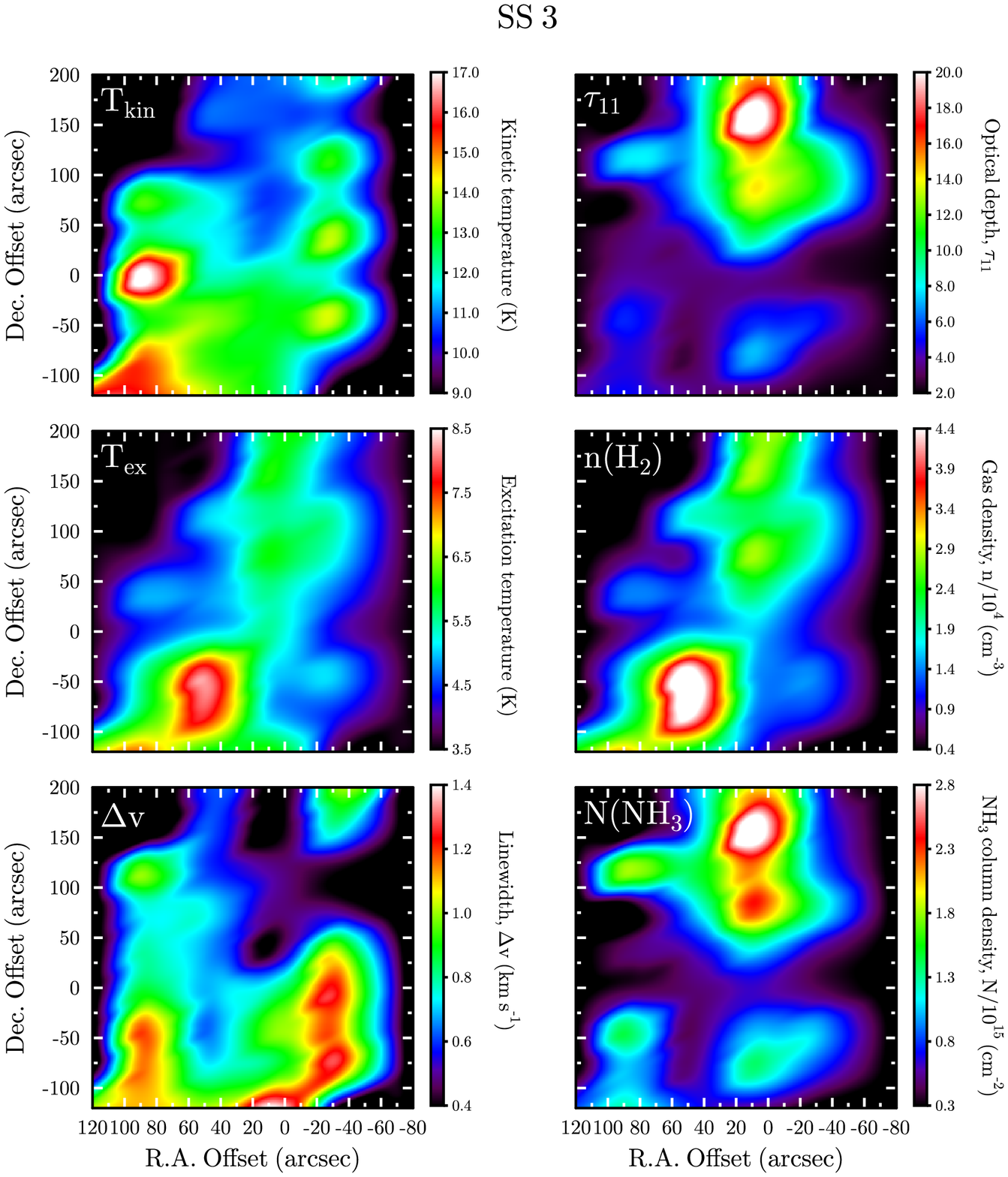,height=18.0cm,width=16.0cm}
\vspace{-2.0cm}
\caption[]{
The spatial distributions of the physical parameters 
measured in \object{SS3} from the ammonia inversion lines \nhhh(1,1) and (2,2).
The corresponding numerical values are listed in Table~\ref{tbl-6}.
}
\label{afg12}
\end{figure*}

The observed spectra of the \nhhh(1,1) and (2,2) transitions detected at the peak positions of ammonia
emission toward each core are shown in Figs.~\ref{afg1}-\ref{afg9}. The measured physical parameters
are listed in Tables~\ref{tbl-3}-\ref{tbl-6}. 
The distributions of the physical parameters for three most abundant cores
are presented in Figs.~\ref{afg10}-\ref{afg12}. 
They demonstrate variations of the kinetic temperature, \Tkin, the excitation temperature, \Tex,
the linewidth, $\Delta v$ (FWHM), the total optical depth, $\tau_{11}$, the gas density,
$n_{{\rm H}_2}$, and the total ammonia column density, $N$(H$_2$), across the mapped area of the core.

\clearpage

\clearpage

\begin{table*}
\centering
\caption{Observed parameters of the \nhhh(1,1) and (2,2) lines and calculated model parameters
for \object{Kawamura~01} and 05, \object{Dobashi 279 P7} and P8, and \object{Dobashi 321 P2} 
}
\label{tbl-3}
\begin{tabular}{cr@{,}l c c c c c c c c c c}
\hline
\hline
\noalign{\smallskip}
\multicolumn{1}{c}{Peak} &
\multicolumn{1}{c}{$\Delta \alpha$} & \multicolumn{1}{c}{$\Delta \delta$} &
\multicolumn{1}{c}{$V_{\scriptscriptstyle\rm LSR}$} & \multicolumn{1}{c}{$\Delta v$} &
\multicolumn{1}{c}{\Tmb} & \multicolumn{1}{c}{\Tex} & \multicolumn{1}{c}{\Trot} & 
\multicolumn{1}{c}{\Tkin} & \multicolumn{1}{c}{$n_{{\rm H}_2}/10^4$} &
\multicolumn{1}{c}{$\tau_{11}$} & \multicolumn{1}{c}{$\tau_{22}$} & 
\multicolumn{1}{c}{$N_{\rm tot}/10^{15}$}\\ 
\multicolumn{1}{c}{Id.} &
\multicolumn{1}{c}{$('')$} & \multicolumn{1}{c}{$('')$} & \multicolumn{1}{c}{(\kms)} &
\multicolumn{1}{c}{(\kms)} & \multicolumn{1}{c}{(K)} & \multicolumn{1}{c}{(K)} &  
\multicolumn{1}{c}{(K)} & \multicolumn{1}{c}{(K)} &
\multicolumn{1}{c}{(\cmm)} & & & \multicolumn{1}{c}{(\cm)} \\ 
\noalign{\smallskip}
\hline
\noalign{\smallskip}
\multicolumn{12}{c}{\object{Ka01}}\\
\noalign{\smallskip}
& 0& 0  & 9.60 & 0.32 & 1.5 & 4.3 & 8.7 & 9.1 & 1.2 & 10.9 & 0.2 & 1.2 \\[-1pt]
$\alpha$ & 0& $-40$& 9.55 & 0.32& 2.5 & 5.5 & 9.2 & 9.6 & 2.5 & 10.5 & 0.3 & 1.3 \\[-1pt]
&+40&$-40$& 9.60 &0.30 & 1.4  &4.8 &$<8.5$ &$\ldots$ &$\ldots$ &4.1 & $<0.1$ & $\geq 0.5$ \\[-1pt]
&+40&$-80$& 9.54 &0.30 & 0.8  &4.1 &$<9.6$ &$\ldots$ &$\ldots$ &3.2 & $<0.1$ &  $\geq 0.3$ \\[-1pt]
&$-40$&0&   9.62 &0.23 & 1.7  &4.7 &$<8.9$ &$\ldots$ &$\ldots$ &8.2 & $<0.2$ &  $\geq 1.0$ \\[-1pt]
&$-40$&$-40$& 9.60 &0.27 & 1.3  &4.3 &$<9.7$ &$\ldots$ &$\ldots$ &6.6 & $<0.2$ &  $\geq 0.5$ \\
\noalign{\smallskip}
\multicolumn{12}{c}{\object{Ka05}}\\
\noalign{\smallskip}
$\alpha$ & $0$&$0$      & 8.40 & 0.37 & 2.5 & 5.6 & 10.2 & 10.7 & 2.3 & 5.6 & 0.3 & 0.8 \\[-1pt]
&$0$&$-40$    & 8.40 & 0.33 & 2.1 & 5.1 & 10.8 & 11.5 & 1.6 & 8.0 & 0.4 & 0.7 \\

\noalign{\smallskip}
\multicolumn{12}{c}{\object{Do279P7}}\\
\noalign{\smallskip}
$\alpha$ & $-80$& 40  & 6.04 & 0.75 & 1.5 & 6.4 & 12.6 & 13.7 & 1.1 & 4.3 & 0.4 & 0.6 \\[-1pt]
&$-40$&$40$ & 6.14 & 0.72 & 1.4 & 4.5 & 13.3 & 14.7 & 1.0 & 3.8 & 0.4 & 0.5  \\[-1pt]
&$-80$&$0$ & 6.59 & 0.55 & 1.4 & 5.1 & 13.5 & 14.9 & 1.4 & 2.4 & 0.3 & 0.3  \\[-1pt]
&$-80$&$80$ & 5.91 & 0.73 & 0.5 & 3.3 & 11.5 & 12.4 & 0.3 & 4.4 & 0.3 & 0.5  \\

\noalign{\smallskip}
\multicolumn{12}{c}{\object{Do279P8}}\\
\noalign{\smallskip}
$\alpha$ & 0& 0  & 8.52 & 0.27 & 2.5 & 5.7 & $<8.0$ &$\ldots$  &$\ldots$  & 8.0 & $<0.1$ & $\geq 1.1$ \\[-1pt]
& 0&$-40$ & 8.57 & 0.30 & 2.4 & 5.7 & $<7.7$ &$\ldots$  &$\ldots$  & 6.9 & $<0.1$ & $\geq 1.2$ \\

\noalign{\smallskip}
\multicolumn{12}{c}{\object{Do321P2}}\\
\noalign{\smallskip}
$\alpha$ & $-80$&$120$  & 6.88 & 0.48 & 3.0 & 6.8 & 11.9 & 12.9 & 3.3 & 4.0 & 0.3 & 0.6 \\[-1pt]
$\beta$ & $-80$&$40$   & 7.05 & 0.43 & 2.7 & 6.7 & 11.4 & 12.2 & 3.4 & 3.5 & 0.2 & 0.5 \\[-1pt]
$\gamma$ & $-120$&$-40$ & 6.93 & 0.38 & 1.7 & 5.5 & $<9.1$ &$\ldots$  &$\ldots$  & 3.0 & $<0.1$ & $\geq 0.5$ \\[-1pt]
$\delta$ & $-80$&$240$  & 7.12 & 0.48 & 1.6 & 4.7 & $<9.7$ &$\ldots$  &$\ldots$  & 4.8 & $<0.1$ & $\geq 0.7$ \\[-1pt]
$\varepsilon$ & $80$&$-40$   & 6.66 & 0.73 & 1.3 & $<5.7$ &$\ldots$  &$\ldots$  &$\ldots$ &$\ldots$  &$\ldots$  &$\ldots$\\

\noalign{\smallskip}
\hline
\noalign{\smallskip}
\multicolumn{13}{l}{{\bf Notes.}\ \Dv\ is the linewidth (FWHM); 
$\tau_{11}$, $\tau_{22}$ are defined in Eqs.~(\ref{Eq2}) and (\ref{Eq3}),   
and $N_{\rm tot}$~--- in Eq.~(\ref{Eq16}).   }
\end{tabular}
\end{table*}

\begin{table*}
\centering
\caption{Observed parameters of the \nhhh(1,1) and (2,2) lines and calculated model parameters
for \object{Do279P6} 
}
\label{tbl-4}
\begin{tabular}{cr@{,}l c c c c c c c c c c}
\hline
\hline
\noalign{\smallskip}
\multicolumn{1}{c}{Peak} &
\multicolumn{1}{c}{$\Delta \alpha$} & \multicolumn{1}{c}{$\Delta \delta$} &
\multicolumn{1}{c}{$V_{\scriptscriptstyle\rm LSR}$} & \multicolumn{1}{c}{$\Delta v$} &
\multicolumn{1}{c}{\Tmb} & \multicolumn{1}{c}{\Tex} & \multicolumn{1}{c}{\Trot} & 
\multicolumn{1}{c}{\Tkin} & \multicolumn{1}{c}{$n_{{\rm H}_2}/10^4$} &
\multicolumn{1}{c}{$\tau_{11}$} & \multicolumn{1}{c}{$\tau_{22}$} & 
\multicolumn{1}{c}{$N_{\rm tot}/10^{15}$}\\ 
\multicolumn{1}{c}{Id.} &
\multicolumn{1}{c}{$('')$} & \multicolumn{1}{c}{$('')$} & \multicolumn{1}{c}{(\kms)} &
\multicolumn{1}{c}{(\kms)} & \multicolumn{1}{c}{(K)} & \multicolumn{1}{c}{(K)} &  
\multicolumn{1}{c}{(K)} & \multicolumn{1}{c}{(K)} &
\multicolumn{1}{c}{(\cmm)} & & & \multicolumn{1}{c}{(\cm)} \\ 
\noalign{\smallskip}
\hline
\noalign{\smallskip}
$\alpha$ & 0& 0         & 7.75 & 0.75 & 2.9 & 5.7 & 12.8 & 14.1 & 1.9 & 5.7 & 0.5 & 1.1 \\[-1pt]
$\beta$ & 40& 40       & 8.04 & 0.63 & 2.1 & 5.3 & 10.8 & 11.5 & 1.8 & 6.0 & 0.3 & 1.1 \\[-1pt]
$\gamma$ & $-160$&$-80$ & 7.69 & 0.96 & 1.8 & 5.8 & 12.9 & 14.2 & 2.0 & 2.2 & 0.2 & 0.5 \\[-1pt]
$\delta$ & $-120$&$-40$ & 7.89 & 0.68 & 1.7 & 5.6 & 10.6 & 11.3 & 2.2 & 2.7 & 0.1 & 0.6 \\[-1pt]
$\varepsilon$ & $80$&$80$    & 7.81 & 0.43 & 1.3 & 4.1 & 9.5  & 9.9  & 0.8 & 10.9& 0.3 & 1.3 \\[-1pt]
&$40$&$0$     & 7.82 & 0.90 & 1.2 & 4.5 & 13.6 & 15.1 & 1.0 & 2.7 & 0.3 & 0.5 \\[-1pt]
&$0$&$40$     & 7.97 & 1.08 & 1.5 & 5.2 & 14.2 & 16.0 & 1.4 & 2.4 & 0.3 & 0.5 \\[-1pt]
&$0$&$-40$    & 7.81 & 0.73 & 1.6 & 4.7 & 14.2 & 15.5 & 1.1 & 4.4 & 0.5 & 0.6 \\[-1pt]
&$-40$&$-40$  & 7.95 & 0.77 & 1.4 & 5.4 & 15.3 & 17.5 & 1.5 & 2.0 & 0.3 & 0.3 \\[-1pt]
&$-40$&$0$    & 7.81 & 0.95 & 1.2 & 4.2 & 11.6 & 12.5 & 0.9 & 4.5 & 0.3 & 0.9 \\[-1pt]
&$-80$&$0$    & 8.08 & 0.38 & 1.2 & 5.2 & 13.2 & 14.5 & 1.4 & 2.4 & 0.2 & 0.2 \\[-1pt]
&$80$&$40$    & 7.78 & 0.40 & 0.9 & 3.9 & 12.2 & 13.2 & 0.6 & 6.0 & 0.4 & 0.4 \\[-1pt]
&$-80$&$-40$  & 8.15 & 0.73 & 1.1 & 5.1 & 12.8 & 14.0 & 1.4 & 1.6 & 0.1 & 0.3 \\[-1pt]
&$-160$&$-40$ & 7.78 & 1.00 & 1.5 & 5.1 & 12.2 & 13.2 & 1.5 & 2.3 & 0.2 & 0.5 \\[-1pt]
&$-200$&$-80$ & 7.96 & 0.83 & 2.2 & 5.5 & 11.9 & 12.9 & 1.9 & 4.2 & 0.3 & 0.9 \\[-1pt]
&$-200$&$-40$ & 7.98 & 0.85 & 1.2 & 4.5 & 10.1 & 10.6 & 1.1 & 3.3 & 0.1 & 0.8 \\[-1pt]
&$-200$&$-120$& 8.19 & 0.68 & 2.2 & 5.3 & 11.3 & 12.1 & 1.7 & 6.7 & 0.4 & 1.2 \\
\noalign{\smallskip}
\hline
\noalign{\smallskip}
\multicolumn{13}{l}{{\bf Notes.}\ \Dv\ is the linewidth (FWHM);
$\tau_{11}$, $\tau_{22}$ are defined in Eqs.~(\ref{Eq2}) and (\ref{Eq3}),   
and $N_{\rm tot}$~--- in Eq.~(\ref{Eq16}).   }
\end{tabular}
\end{table*}

\begin{table*}
\centering
\caption{Observed parameters of the \nhhh(1,1) and (2,2) lines and calculated model parameters
for \object{Do279P12} 
}
\label{tbl-5}
\begin{tabular}{cr@{,}l c c c c c c c c c c}
\hline
\hline
\noalign{\smallskip}
\multicolumn{1}{c}{Peak} &
\multicolumn{1}{c}{$\Delta \alpha$} & \multicolumn{1}{c}{$\Delta \delta$} &
\multicolumn{1}{c}{$V_{\scriptscriptstyle\rm LSR}$} & \multicolumn{1}{c}{$\Delta v$} &
\multicolumn{1}{c}{\Tmb} & \multicolumn{1}{c}{\Tex} & \multicolumn{1}{c}{\Trot} & 
\multicolumn{1}{c}{\Tkin} & \multicolumn{1}{c}{$n_{{\rm H}_2}/10^4$} &
\multicolumn{1}{c}{$\tau_{11}$} & \multicolumn{1}{c}{$\tau_{22}$} & 
\multicolumn{1}{c}{$N_{\rm tot}/10^{15}$}\\ 
\multicolumn{1}{c}{Id.} &
\multicolumn{1}{c}{$('')$} & \multicolumn{1}{c}{$('')$} & \multicolumn{1}{c}{(\kms)} &
\multicolumn{1}{c}{(\kms)} & \multicolumn{1}{c}{(K)} & \multicolumn{1}{c}{(K)} &  
\multicolumn{1}{c}{(K)} & \multicolumn{1}{c}{(K)} &
\multicolumn{1}{c}{(\cmm)} & & & \multicolumn{1}{c}{(\cm)} \\ 
\noalign{\smallskip}
\hline
\noalign{\smallskip}
$\alpha$ & $160$&$-40$  & 7.59  & 0.90 & 3.2 & 7.6 & 15.6 & 17.8 & 3.2 & 2.8 & 0.4 & 0.7 \\[-1pt]
$\beta$ & $200$&$-120^a$ &A 6.94 & 0.75 & 3.1 & 7.4 & 13.5 & 14.9 & 3.5 & 3.9 & 0.4 & 0.8 \\[-1pt]
&\multicolumn{2}{c}{ }&B 8.21 & 0.73 & 1.2  & 5.3 & 14.1  & 15.8 & 1.5 & 1.7 & 0.2  & 0.3 \\[-1pt]
&$200$&$-200^a$ &A 7.26 & 0.62 & 2.3 & 7.5 & 11.7 & 12.6 & 4.5 & 1.9 & 0.1 & 0.4 \\[-1pt]
&\multicolumn{2}{c}{ }&B 8.58 & 0.77 & 2.0  & 5.8 & 12.6  & 13.7 & 2.1 & 2.8 & 0.2  & 0.6 \\[-1pt]
$\gamma$ & 40& 40       & 8.29  & 0.90 & 2.9 & 6.4 & 12.2 & 13.2 & 2.8 & 4.1 & 0.3 & 1.1 \\[-1pt]
&0& 0         & 8.23  & 1.22 & 1.3 & 4.9 & 14.6 & 16.5 & 1.1 & 2.3 & 0.3 & 0.5 \\[-1pt]
&40& 0        & 8.42  & 1.15 & 1.7 & 5.1 & 12.8 & 14.0 & 1.4 & 3.0 & 0.3 & 0.8 \\[-1pt]
&80& 0$^a$    &A 7.89 & 0.68 & 1.3 & 6.9 & 16.4 & 19.0 & 2.5 & 1.0 & 0.2 & 0.2 \\[-1pt]
&\multicolumn{2}{c}{ }&B 8.92 & 0.83 & 0.6  & 5.1  & 14.1  & 15.8  & 1.3 & 0.7 & 0.1  & 0.1 \\[-1pt]
&0& 80        & 8.44  & 0.98 & 1.8 & 5.3 & 15.8 & 18.2 & 1.4 & 3.1 & 0.5 & 0.6 \\[-1pt]
&40& 80       & 8.35  & 0.77 & 2.6 & 6.0 & 12.9 & 14.1 & 2.2 & 4.5 & 0.4 & 0.9 \\[-1pt]
&40& 120      & 8.42  & 0.68 & 2.8 & 6.5 & 12.8 & 14.0 & 2.6 & 4.0 & 0.4 & 0.8 \\[-1pt]
&40&$-40$     & 7.91  & 0.77 & 1.3 & 4.9 & 13.8 & 15.4 & 1.2 & 2.4 & 0.3 & 0.4 \\[-1pt]
&0&$120$      & 8.56  & 1.03 & 2.0 & 5.6 & 14.7 & 16.6 & 1.7 & 3.0 & 0.4 & 0.7 \\[-1pt]
&80&$80$      & 8.19  & 0.97 & 1.3 & 4.3 & 13.9 & 15.4 & 0.8 & 4.3 & 0.5 & 0.7 \\[-1pt]
&80&$40$      & 8.51  & 0.80 & 2.2 & 5.5 & 11.8 & 12.8 & 1.9 & 4.5 & 0.3 & 0.9 \\[-1pt]
&120&$-40$    & 8.16  & 0.73 & 1.6 & 5.0 & 13.8 & 15.4 & 1.2 & 3.6 & 0.4 & 0.5 \\[-1pt]
&120&$-80$    & 8.15  & 1.17 & 1.8 & 7.0 & 12.3 & 13.4 & 3.4 & 1.3 & 0.1 & 0.5 \\[-1pt]
&160&$-80$    & 7.67  & 1.25 & 2.4 & 6.4 & 14.0 & 15.6 & 2.4 & 2.6 & 0.3 & 0.8 \\[-1pt]
&160&$-120^a$   &A 7.02 & 0.68 & 1.6 & 7.0 & 14.5 & 16.2 & 2.9 & 1.3 & 0.2 & 0.2 \\[-1pt]
&\multicolumn{2}{c}{ }  &B 8.39 & 0.90 & 1.5 & 5.6 & 14.8 & 16.8 & 1.6 & 1.9 & 0.3 & 0.4 \\[-1pt]
&160&$-160^a$   &A 7.40 & 1.5 & 1.0 &$\ldots$& $\ldots$& $\ldots$& $\ldots$ & 0.3 &$\ldots$  & $\ldots$ \\[-1pt]
&\multicolumn{2}{c}{ }&B 8.64 & 0.7 & 1.3 &$\ldots$  &$\ldots$  &$\ldots$  &$\ldots$ & 1.3 &$\ldots$  &$\ldots$  \\[-1pt]
&160&$-200^a$   &A 7.18 & 0.3 & 1.1 &$\ldots$  &$\ldots$  &$\ldots$  &$\ldots$  & 1.0 &$\ldots$  &$\ldots$  \\[-1pt]
&\multicolumn{2}{c}{ }&B 8.59 & 1.0 & 1.0 &$\ldots$  &$\ldots$  & $\ldots$ &$\ldots$ & 1.4 &$\ldots$  &$\ldots$  \\[-1pt]
&240&$40$     &  7.80 & 0.73 & 1.6 & 5.6 & 13.6 & 15.1 & 1.7 & 2.3 & 0.2 & 0.4 \\[-1pt]
&200&$-40$    &  7.88 & 0.87 & 2.1 & 10.2 & 15.2 & 17.3 & 7.0 & 0.8 & 0.1 & 0.3 \\[-1pt]
&200&$-80$    &  7.06 & 1.28 & 1.9 &  6.5 & 14.1 & 15.7 & 2.5 & 1.5 & 0.2 & 0.5 \\[-1pt]
&200&$-160^a$   &A 7.26 & 0.62 & 2.3 &  7.5 & 11.7 & 12.6 & 4.5 & 1.8 & 0.1 & 0.4 \\[-1pt]
&\multicolumn{2}{c}{ } &B 8.58 & 0.77 & 2.0 &  5.8 & 12.6 & 13.7 & 2.1 & 2.8 & 0.2 & 0.6 \\[-1pt]
&200&$-240$   & 8.29  & 0.83 & 1.8 &  5.2 & 12.8 & 14.1 & 1.5 & 3.4 & 0.3 & 0.6 \\[-1pt]
&200&$-280$   & 8.42  & 0.60 & 1.9 &  5.6 & 11.3 & 12.1 & 1.9 & 3.5 & 0.2 & 0.6 \\
\noalign{\smallskip}
\hline
\noalign{\smallskip}
\multicolumn{13}{l}{{\bf Notes.}\ $^a$Partly resolved velocity components  
marked by letters A and B are shown in Fig.~\ref{afg4}.  }\\
\multicolumn{13}{l}{\Dv\ is the linewidth (FWHM);
$\tau_{11}$, $\tau_{22}$ are defined in Eqs.~(\ref{Eq2}) and (\ref{Eq3}),   
and $N_{\rm tot}$~--- in Eq.~(\ref{Eq16}).   }
\end{tabular}
\end{table*}

\begin{table*}
\centering
\caption{Observed parameters of the \nhhh(1,1) and (2,2) lines and calculated model parameters
for \object{SS3} 
}
\label{tbl-6}
\begin{tabular}{cr@{,}l c c c c c c c c c c}
\hline
\hline
\noalign{\smallskip}
\multicolumn{1}{c}{Peak} &
\multicolumn{1}{c}{$\Delta \alpha$} & \multicolumn{1}{c}{$\Delta \delta$} &
\multicolumn{1}{c}{$V_{\scriptscriptstyle\rm LSR}$} & \multicolumn{1}{c}{$\Delta v$} &
\multicolumn{1}{c}{\Tmb} & \multicolumn{1}{c}{\Tex} & \multicolumn{1}{c}{\Trot} & 
\multicolumn{1}{c}{\Tkin} & \multicolumn{1}{c}{$n_{{\rm H}_2}/10^4$} &
\multicolumn{1}{c}{$\tau_{11}$} & \multicolumn{1}{c}{$\tau_{22}$} & 
\multicolumn{1}{c}{$N_{\rm tot}/10^{15}$}\\ 
\multicolumn{1}{c}{Id.} &
\multicolumn{1}{c}{$('')$} & \multicolumn{1}{c}{$('')$} & \multicolumn{1}{c}{(\kms)} &
\multicolumn{1}{c}{(\kms)} & \multicolumn{1}{c}{(K)} & \multicolumn{1}{c}{(K)} &  
\multicolumn{1}{c}{(K)} & \multicolumn{1}{c}{(K)} &
\multicolumn{1}{c}{(\cmm)} & & & \multicolumn{1}{c}{(\cm)} \\ 
\noalign{\smallskip}
\hline
\noalign{\smallskip}
&$0$&$0$     & 7.46  & 0.83 & 1.6 & 5.3 & 11.4 & 12.2 & 1.5 & 2.8 & 0.2 & 0.6 \\[-1pt]
&$0$&$40$    & 7.60  & 0.47 & 2.5 & 5.4 & 10.7 & 11.3 & 2.0 & 8.3 & 0.4 & 1.1 \\[-1pt]
$\gamma$ &$0$&$80$    & 7.67  & 0.48 & 3.1 & 5.9 & 10.2 & 10.7 & 2.7 & 13.2 & 0.5 & 2.2 \\[-1pt]
&$0$&$120$   & 7.66  & 0.48 & 2.7 & 5.5 & 10.5 & 11.2 & 2.1 & 13.8 & 0.6 & 2.0 \\[-1pt]
$\beta$ & $0$&$160$   & 7.62  & 0.40 & 3.1 & 5.9 & 10.1 & 10.7 & 2.7 & 20.0 & 0.7 & 2.8 \\[-1pt]
&$0$&$200$   & 7.52  & 0.40 & 3.2 & 6.0 & 10.5 & 11.1 & 2.8 & 14.5 & 0.6 & 2.0 \\[-1pt]
&$0$&$-40$   & 7.80  & 1.00 & 1.9 & 5.0 & 12.0 & 13.0 & 1.4 & 4.3 & 0.3 & 1.0 \\[-1pt]
&$0$&$-80$   & 7.89  & 0.90 & 1.9 & 4.8 & 11.6 & 12.4 & 1.3 & 6.0 & 0.4 & 1.3 \\[-1pt]
&$0$&$-120$  & 7.27  & 1.37 & 0.8 & 5.9 & 11.4 & 12.2 & 0.9 & 1.7 & 0.1 & 0.5 \\[-1pt]
&$120$&$-120$  & 7.67  & 1.00 & 2.8 & 6.5 & 14.1 & 15.7 & 2.4 & 3.1 & 0.4 & 0.8 \\[-1pt]
&$40$&$0$    & 7.45  & 0.70 & 1.5 & 5.1 & 11.2 & 12.0 & 1.6 & 2.6 & 0.1 & 0.5 \\[-1pt]
&$40$&$40$   & 7.54  & 0.64 & 1.7 & 4.9 & 10.9 & 11.6 & 1.4 & 4.1 & 0.2 & 0.7 \\[-1pt]
&$40$&$80$   & 7.81  & 0.70 & 1.8 & 4.7 & 10.9 & 11.6 & 1.2 & 7.2 & 0.4 & 1.2 \\[-1pt]
&$40$&$120$  & 7.80  & 0.58 & 2.2 & 5.1 & 9.6 & 10.0 & 1.9 & 7.5 & 0.2 & 1.5 \\[-1pt]
&$40$&$160$  & 7.66  & 0.60 & 1.2 & 4.0 & 10.3 & 10.9 & 0.8 & 7.7 & 0.3 & 1.1 \\[-1pt]
&$40$&$200$  & 7.62  & 0.62 & 1.0 & 3.8 & 9.9 & 10.4 & 0.6 & 7.2 & 0.2 & 1.1 \\[-1pt]
&$40$&$-40^a$ &A 7.85 & 0.67 & 2.9 & 7.6 & 12.5 & 13.6 & 4.2 & 2.4 & 0.2 & 0.5 \\[-1pt]
&\multicolumn{2}{c}{ }&B 6.86 & 0.57 & 1.1  & 5.3 & 11.3  & 12.1 & 1.8 & 1.5 & 0.1  & 0.2 \\[-1pt]
&$40$&$-80^a$ &A 7.88 & 0.75 & 2.9 & 7.5 & 12.1 & 13.1 & 4.2 & 2.4 & 0.2 & 0.6 \\[-1pt]
&\multicolumn{2}{c}{ }&B 6.85 & 0.57 & 1.2  & 5.4 & 12.4  & 13.5 & 1.7 & 1.7 & 0.1  & 0.2 \\[-1pt]
&$40$&$-120$ & 7.28  & 1.03 & 2.2 & 5.9 & 12.6 & 13.8 & 2.1 & 2.6 & 0.2 & 0.7 \\[-1pt]
&$80$&$0$    & 7.34  & 0.85 & 1.2 & 4.5 & 14.8 & 16.8 & 0.9 & 2.9 & 0.4 & 0.4 \\[-1pt]
&$80$&$40$   & 7.37  & 0.78 & 1.4 & 4.9 & 11.6 & 12.4 & 1.4 & 2.6 & 0.2 & 0.5 \\[-1pt]
&$80$&$80$   & 7.56  & 0.75 & 0.7 & 3.9 & 12.1 & 13.1 & 0.7 & 2.2 & 0.2 & 0.3 \\[-1pt]
&$80$&$120$  & 7.53  & 0.93 & 0.5 & 3.3 & 8.6 & 8.9 & 0.3 & 7.2 & 0.1 & 1.8 \\[-1pt]
&$80$&$-40$  & 7.36  & 1.13 & 2.6 & 5.8 & 12.2 & 13.2 & 2.1 & 4.3 & 0.3 & 1.3 \\[-1pt]
&$80$&$-80$  & 7.68  & 1.12 & 2.7 & 6.1 & 13.5 & 14.9 & 2.2 & 3.6 & 0.4 & 1.0 \\[-1pt]
&$80$&$-120$ & 7.57  & 1.10 & 3.6 & 7.1 & 13.9 & 15.4 & 3.1 & 4.0 & 0.4 & 1.2 \\[-1pt]
$\alpha$ & $80$&$-160$ & 7.39  & 1.13 & 5.0 & 8.4 & 14.7 & 16.6 & 4.4 & 4.7 & 0.6 & 1.7 \\[-1pt]
&$-40$&$0$   & 7.46  & 1.12 & 0.7 & 4.1 & 10.5 & 11.1 & 0.8 & 1.6 & 0.1 & 0.4 \\[-1pt]
&$-40$&$40$  & 7.51  & 0.92 & 1.2 & 4.5 & 12.2 & 13.2 & 1.0 & 2.7 & 0.2 & 0.5 \\[-1pt]
&$-40$&$80$  & 7.57  & 0.45 & 2.1 & 5.0 & 10.7 & 11.4 & 1.5 & 8.6 & 0.4 & 1.1 \\[-1pt]
&$-40$&$120$ & 7.55  & 0.40 & 2.2 & 5.1 & 11.5 & 12.4 & 1.6 & 8.2 & 0.5 & 0.8 \\[-1pt]
&$-40$&$160$ & 7.56  & 0.73 & 1.2 & 4.4 & 9.8 & 10.3 & 1.1 & 3.4 & 0.1 & 0.7 \\[-1pt]
&$-40$&$200$ & 7.31  & 1.00 & 1.3 & 4.7 & 11.3 & 12.1 & 1.2 & 2.5 & 0.1 & 0.6 \\[-1pt]
&$-40$&$-40$ & 7.76  & 1.05 & 1.7 & 4.8 & 12.2 & 13.2 & 1.2 & 3.8 & 0.3 & 0.9 \\[-1pt]
&$-40$&$-80$ & 7.76  & 1.13 & 1.0 & 4.3 & 9.8 & 10.3 & 1.0 & 2.2 & 0.1 & 0.7 \\
\noalign{\smallskip}
\hline
\noalign{\smallskip}
\multicolumn{13}{l}{{\bf Notes.}\ $^a$Partly resolved velocity components  
marked by letters A and B are shown in Fig.~\ref{afg8}.  }\\
\multicolumn{13}{l}{\Dv\ is the linewidth (FWHM);
$\tau_{11}$, $\tau_{22}$ are defined in Eqs.~(\ref{Eq2}) and (\ref{Eq3}),   
and $N_{\rm tot}$~--- in Eq.~(\ref{Eq16}).   }
\end{tabular}
\end{table*}

\end{appendix} 

\end{document}